\pdfoutput=1
\RequirePackage{fix-cm}
\documentclass[smallcondensed]{svjour3}     
\smartqed  

\usepackage{amsmath}
\usepackage{graphicx}
\usepackage{booktabs}
\usepackage{amssymb}
\usepackage{latexsym}	
\usepackage{array}		
\usepackage{multirow}
\usepackage{subfigure}
\usepackage{algorithm}
\usepackage{algpseudocode}
\usepackage{threeparttable}
\usepackage{paralist}   
\usepackage{xspace}
\usepackage{color}
\usepackage{xcolor}
\usepackage{adjustbox}
\usepackage{balance}
\usepackage{wasysym}
\usepackage{rotating}   
\usepackage{listings}
\usepackage{caption}    

\usepackage{hyperref}
\usepackage{url}            

\newcommand{\red}[1]{\textcolor[rgb]{0.00,0.00,0.00}{#1}}
\newcommand{\blue}[1]{\textcolor[rgb]{0.00,0.00,0.00}{#1}}

\definecolor{mygreen}{rgb}{0,0.6,0}
\definecolor{mygray}{rgb}{0.5,0.5,0.5}
\definecolor{mymauve}{rgb}{0.58,0,0.82}
\definecolor{darkblue}{rgb}{0.0,0.0,0.6}
\definecolor{maroon}{RGB}{102, 0, 0}
\definecolor{Maroon}{cmyk}{0,0.87,0.68,0.32}
\definecolor{darkred}{RGB}{139, 0, 0}

\lstdefinelanguage{XML}
{
  basicstyle=\ttfamily\small,   
  morestring=[b]",
  moredelim=[s][\color{darkblue}]{<}{\ },
  moredelim=[s][\color{darkblue}]{</}{>},
  moredelim=[l][\color{darkblue}]{/>},
  moredelim=[l][\color{darkblue}]{>},
  morecomment=[s]{<?}{?>},
  morecomment=[s]{<!--}{-->},
  stringstyle=\color{darkred},
  identifierstyle=\color{mymauve}
}

\lstdefinestyle{customJava}{
  breaklines=true,
  keepspaces=true,
  frame=single,
  language=Java,
  showstringspaces=false,
  basicstyle=\footnotesize\ttfamily,
  keywordstyle=\color{blue},
  otherkeywords={+, getIntent},
  numbers=left,
  numbersep=5pt,
  numberstyle=\scriptsize\color{black},
  rulecolor=\color{black},
  stepnumber=1,
  tabsize=2,
  commentstyle=\itshape\color{green!40!black},
  stringstyle=\color{orange},
  emph=[1]  
  {
        do,
        try,
        new,
        catch,
        while,
        SecReceiver,
        SecService,
        SecActivity,
        SecSink,
  },
  emphstyle=[1]{\color{darkred}},
  emph=[2]  
  {
        @Override,
        SDK\_INT,
        LOLLIPOP,
        builder,
        PkgName
  },
  emphstyle=[2]{\color{mymauve}},
}

\newif\ifACM
\ACMfalse 

\ifACM
\newcommand{\myfig}{Figure}
\else
\newcommand{\myfig}{Fig.}
\fi

\ifACM
\newcommand{\mysec}{Section~}
\else
\newcommand{\mysec}{Sec.~}
\fi

\newcommand{\minSDK}{\texttt{minSdkVersion}\xspace}
\newcommand{\aimSDK}{\texttt{targetSdkVersion}\xspace}
\newcommand{\maxSDK}{\texttt{maxSdkVersion}\xspace}
\newcommand{\DSDK}{\texttt{DSDK}\xspace}
\newcommand{\lagSDK}{\texttt{lagSdkVersion}\xspace}

\newcommand{\minLevel}{\texttt{minLevel}\xspace}
\newcommand{\maxLevel}{\texttt{maxLevel}\xspace}

\newcommand{\minOverNum}{\texttt{minOverNum}\xspace}

\newcommand{\aapt}{\texttt{aapt}\xspace}
\newcommand{\apktool}{\texttt{apktool}\xspace}
\newcommand{\dexdump}{\texttt{dexdump}\xspace}

\journalname{Empirical Software Engineering}

\begin{document}

\title{Scalable Online Vetting of Android Apps for Measuring Declared SDK Versions and Their Consistency with API Calls
}

\titlerunning{Measuring Declared SDK Versions and Their Consistency with API Calls}        

\author{Daoyuan Wu \and
        Debin Gao \and
        David Lo
}


\institute{Daoyuan Wu \at
           \email{dywu@ie.cuhk.edu.hk}           
           \at Department of Information Engineering, The Chinese University of Hong Kong, Hong Kong.
           \and
           Debin Gao \at
           \email{dbgao@smu.edu.sg}           
           \at School of Information Systems, Singapore Management University, Singapore.
           \and
           David Lo \at
           \email{davidlo@smu.edu.sg}           
           \at School of Information Systems, Singapore Management University, Singapore.
}

\date{Received: March 2019 / Accepted: date}

\maketitle

\begin{abstract}

Android has been the most popular smartphone system with multiple platform versions active in the market.
To manage the application's compatibility with one or more platform versions, Android allows apps to declare the supported platform SDK versions in their manifest files.
In this paper, we \red{conduct} a systematic study of this modern software mechanism.
Our objective is to measure the current practice of declared SDK versions (which we term as \DSDK versions afterwards) in real apps, and the (in)consistency between \DSDK versions and their host apps' API calls.
To successfully analyze a modern dataset of 22,687 popular apps (with an average app size of 25MB), we design a scalable approach that operates on the Android bytecode level and employs a lightweight bytecode search for app analysis.
This approach achieves a good performance suitable for online vetting in app markets, requiring only around 5 seconds to process an app on average.
Besides shedding light on the characteristics of \DSDK in the wild, our study quantitatively measures two side effects of inappropriate \DSDK versions: (i) \red{around 35\%} apps under-set the minimum \DSDK versions and could incur runtime crashes, but fortunately, only 11.3\% apps could crash on Android 6.0 and above; (ii) around 2\% apps, due to under-claiming the targeted \DSDK versions, are potentially exploitable by remote code execution, and half of them invoke the vulnerable API via embedded third-party libraries.
These results indicate the importance and difficulty of declaring correct \DSDK, and our work can help developers fulfill this goal.


\end{abstract}

\section{Introduction}
\label{sec:intro}

\red{In} recent years\red{,} \red{we} have witnessed the extraordinary success of Android, a smartphone operating system owned by Google.
At the end of 2013, Android became the best-selling phone and tablet OS.
As of 2015, Android evolved into the largest installed base of all operating systems.
Over these years, Android keeps leading the global smartphone market share at over 80\%~\cite{mobileOSshare}.
Along with the fast-evolving Android, its fragmentation problem becomes more and more serious.
Although new devices ship with the recent Android versions, there are still huge amounts of existing devices running \red{old versions of Android}~\cite{dashboards}.

To better manage the application's compatibility \red{across} multiple platform versions, Android allows apps to declare the supported platform SDK versions in their \red{so-called ``}manifest\red{'' app configuration} files \red{(manifest afterwards)}.
We term these declared SDK versions as \DSDK versions.
The \DSDK mechanism is a modern software mechanism with which, to the best of our knowledge, few systems are equipped until Android.
Nevertheless, it receives little attention so far, and few understandings are known about the effectiveness of the \DSDK mechanism.

In this paper, we aim to \red{conduct} a systematic study on the Android \DSDK mechanism.
Specifically, our objective is to measure the current practice of \DSDK versions in real apps, and the (in)consistency between \DSDK versions and their host apps' API calls.
To make our measurement results representative, we select popular apps with at least one million installs each on Google Play as the dataset.
More specifically, we have collected a large-scale dataset with 22,687 popular apps (570.8GB in total, with an average app size of 25MB), which covers 90.2\% of all such apps (both free and paid ones) available on Google Play. 
Furthermore, our study utilizes the latest Android API evolution and covers all 28 versions of Android SDKs or API levels\footnote{The latest Android version at the time of our writing is Android 9 (API level 28).}. 

After \red{collecting} the dataset and building the API-SDK mapping, we perform a systematic \DSDK and API call analysis of each individual app.
\red{We design our approach} \blue{scalable and} robust so that it can be readily deployed by online app markets (e.g., Google Play) to timely notify developers of the \DSDK inconsistency in their apps.
Given this objective, dataflow-based analysis is not suitable because existing Android dataflow analyses (notably FlowDroid~\cite{FlowDroid14} and Amandroid~\cite{Amandroid14}) are expensive even when analyzing medium-size\red{d} apps, e.g., requiring $\sim$4 minutes for the 8MB Nextcloud app\footnote{\url{https://f-droid.org/en/packages/com.nextcloud.client/}}~\cite{IctApiFinder18}.
Moreover, they need to first transform or decompile Android app bytecode into an intermediate representation (\blue{e.g., Soot Jimple or Java bytecode}), the process of which is not fully accurate~\cite{RetargetDex12} and often leaves some apps unanalyzable in many previous studies~\cite{AppContext15}~\cite{MudFlow15}~\cite{MaMaDroid17}~\cite{HSOMiner17}.

In our approach, we thus operate on the original Android bytecode level and employ \red{a} lightweight \textit{bytecode search} for app analysis.
Specifically, we retrieve \DSDK versions and API calls \textit{directly} from each app without decoding the manifest file \blue{and without} \blue{transforming} app bytecode, which enables robust processing of all 22,687 popular apps.
We also handle multidex~\cite{multidex}, a special Android bytecode \blue{mechanism} \red{that is} often skipped by prior works but common in modern apps --- 5,008 apps in our dataset split their bytecodes into multiple files.
With the correctly extracted app bytecodes, we then search these bytecode texts to obtain valid API calls that are not guarded by \texttt{VERSION.SDK\_INT} checking (developers can use such \texttt{if} statements to invoke an API only in certain Android platforms) and \red{are} also not in uninvoked third-party libraries.
In this way, our approach preserves the scalability and \red{makes itself} suitable for online vetting: the median and average time for analyzing an app in our dataset is only 4.75s and 5.39s, \red{respectively}.

\blue{Theoretically, our lightweight approach is less accurate than dataflow-based approaches.
This is because we did not perform (the expensive) flow tracking, and false positives certainly appear.
Fortunately, this limitation would not affect the real usage of our approach, since in our objective, the approach is used by online app markets for checking apps uploaded by developers.
In other words, we can ask developers to manually check the inconsistency warnings in their apps.
Moreover, the manual effort required in such checking is also limited --- around 80\% apps have fewer than ten potentially inconsistent API calls each.
This indicates that the number of inconsistency warnings per app reported by our bytecode search is well manageable for developers to perform a one-time manual check.
It is worth noting that this paper is not for bug detection; instead, we aim for a comprehensive study on the current \DSDK practice and its potential impacts.
By employing a lightweight yet conservative approach, we can maximize the coverage of valid code and thus minimize false negatives (the dataflow tracking is sometimes too tight and could fail to process complex implicit flows, e.g., as high as 13 different kinds of implicit flows missed in FlowDroid according to a systematic assessment recently~\cite{uSE18}).}

\blue{In a nutshell,} our study sheds light on the current \DSDK practice \red{of} app developers and quantitatively measures two side effects caused by the inconsistency between \DSDK versions (configured by the app developers in the manifest file) and API calls (made by the app during its execution).
Specifically, the compatibility effect occurs when a minimum \DSDK version is set too low \red{so} that certain APIs do not even exist in \red{the} corresponding low\red{er} versions of Android platforms.
The consequence of such compatibility effect can cause runtime crashes.
Additionally, the security effect could also happen when a target \DSDK version is outdated (i.e., a lower version \red{of APIs will be} used \red{even when a} device runs on later versions of Android), causing that a vulnerable API is still rendered by the underlying system when the app runs on higher versions of Android.
\red{Next}, we present our \red{three sets of} measurement results on \DSDK versions and their inconsistency with API calls.
\blue{Note that due to the conservative nature of our approach, the measurement results reported in this paper represent an upper bound of all potential \DSDK problems (under the condition that common analysis difficulties, such as native code, are not considered).}

Firstly, our measurement \red{reveals} some interesting characteristics of declared SDK versions in the wild. 
Specifically, nearly all apps define the \minSDK attribute, but 4.76\% apps still do not claim the \aimSDK attribute (in our dataset \red{collected} in \red{late} 2018).
\red{Fortunately}, this percentage has significantly dropped from 16.54\% in 2015, \red{which} indicates that \DSDK attributes nowadays are more widely adopted in modern apps.
We further find that the minimal platform version most apps support nowadays is Android 4.1, whereas the most popular targeted platform version is Android 8.0.
The median version difference between \aimSDK and \minSDK also increases from 8 (in our last analysis in 2015) to 9 (currently in the 2018 dataset).

Secondly, in terms of compatibility inconsistency, we find that \red{around 35\%} apps under-set the \minSDK value, causing them to crash when running on lower versions of Android platforms.
Fortunately, only 11.3\% apps could crash on Android 6.0 and above.
We also show that by employing bytecode search for \texttt{SDK\_INT} checking, our approach can reduce 17.3\% false positives of compatibility inconsistency results.
A detailed analysis of \red{the} Android APIs \red{incurring} compatibility inconsistency further reveals that some API classes, such as view, webkit, and system manager related classes, are commonly misused.

Thirdly, our analysis of security inconsistency shows that around 2\% apps set an outdated \aimSDK attribute \red{and also invoke} a \red{dangerous} WebView API, making them\red{selves} exploitable by remote code execution.
In particular, around half of these vulnerable apps invoke the vulnerable \texttt{addJavascriptInterface()} API \red{only} because of their embedded third-party libraries.
\red{Additionally}, our bytecode search of the \texttt{addJavascriptInterface()} invocation also helps reduce 12.2\% false positives.

To summarize, we highlight the contributions of this paper as follows:
\begin{itemize}
\item (\textit{New problem})
  To the best of our knowledge, we are the first to conduct a systematic study on \DSDK, a modern software mechanism that allows apps to declare the supported platform SDK versions.
  We also give the first demystification of the \DSDK mechanism and its two side effects on compatibility and security.
  In particular, our \red{preliminary} conference version of this work~\cite{WASA17} has \red{motivated} several recent follow-up works~\cite{CiD18}~\cite{IctApiFinder18} on bug detection.

\item (\textit{Novel approach})
  We propose a robust and scalable approach that operates directly on the original bytecode level and leverages lightweight bytecode search to timely notify developers of the \DSDK inconsistency in their apps.
  The evaluation using 22,687 popular apps (with an average app size as large as 25MB) shows that our approach achieves a good performance suitable for online app vetting, requiring only around 5 seconds to process an app on average.
  
\item (\textit{New findings})
  Our measurement study obtains three major new findings, including (i) 4.76\% apps still do not claim the \aimSDK attribute, although this percentage has significantly dropped from 2015 to 2018, (ii) \red{around 35\%} apps under-set the minimum \DSDK versions and could incur runtime crashes, but fortunately, only 11.3\% apps could crash on Android 6.0 and above, and (iii) around 2\% apps, due to under-claiming the targeted \DSDK versions, are potentially exploitable by remote code execution, and half of them actually invoke the vulnerable API via embedded third-party libraries.
\end{itemize}

In this journal article, we extend our preliminary conference version~\cite{WASA17} from the following perspectives:
(1) We integrate a lightweight bytecode search into our approach so that it can be deployed by online app markets to timely notify developers of the DSDK inconsistency in their apps. We also add support for multidex-based apps and enhance the detection of uninvoked third-party libraries.
(2) We evolve our dataset from an old set of 23,125 random apps in 2015 to a recent set of 22,687 popular apps in November 2018. We also find a lightweight way to build the latest API-SDK mapping.
(3) By running experiments using the improved approach and dataset, we obtain more representative results and compare some of our new findings with the previous ones.


\section{Demystifying Declared SDK Versions and Their Two Side Effects}
\label{sec:backg}

In this section, we first demystify declared platform SDK versions in Android apps, and then explain their two side effects if inappropriate \DSDK versions are used.
Note that \DSDK is different from the typical compilation SDK, which is only for compiling apps while \DSDK is mainly for interpreting run-time API behaviors.

\subsection{Declared SDK Versions in Android Apps}
\label{sec:declared}

Listing~\ref{lst:sdkversion} illustrates how to declare supported platform SDK versions in Android apps by defining the \texttt{<uses-sdk>} element~\cite{uses-sdk} in apps' manifest files (i.e., \texttt{AndroidManifest.xml}~\cite{manifest}).
These \DSDK versions are for the runtime Android system to check apps' compatibility, which is different from the compiling-time SDK for compiling source codes.
The value of each \DSDK version is an integer, which represents the API level of the corresponding SDK.
For example, if a developer wants to declare Android SDK version 5.0, she can set its value to 21.
Since each API level has a precise mapping of the corresponding SDK version~\cite{AndroidVersion}, we do not use another term, \textit{declared API level}, to represent the same meaning of \DSDK throughout this paper.

\begin{lstlisting}[
language=XML,
basicstyle=\ttfamily,
label={lst:sdkversion},
caption={The syntax for declaring platform SDK versions in Android apps.}
]
<uses-sdk android:minSdkVersion="integer"
        android:targetSdkVersion="integer"
        android:maxSdkVersion="integer" />
\end{lstlisting}

We explain the three \DSDK attributes as follows: 
\begin{itemize}
  \item The \minSDK integer specifies the minimum platform API level required for an app to run. The Android system refuses to install an app if its \minSDK value is greater than the system's API level. Note that if an app does not declare this attribute, the system by default assigns the value of ``1'', which means that the app can be installed in all versions of Android.

  \item The \aimSDK integer designates the platform API level that an app targets at. An important \textit{implication} of this attribute is that Android adopts backward-compatible API behaviors of the declared target SDK version, even when an app is running on a higher version of the Android platform. Android makes such compromised design because it aims to guarantee the same app behaviors \red{as expected by developers}, even when apps run on newer platforms. It is worth noting that if this attribute is not set, the \minSDK is used.

  \item The \maxSDK integer specifies the maximum platform API level on which an app can run. However, this attribute is \textit{not} recommended and already deprecated since Android 2.1 (API level 7). Modern Android no longer checks or enforces this attribute during the app installation or re-validation. The only effect is that Google Play continues to use this attribute as a filter when it presents users a list of applications available for downloading. Not\red{e} that if this attribute is not set, it implies no restriction on the maximum platform API level.
\end{itemize}

\subsection{Two Side Effects of Inappropriate DSDK Versions}
\label{sec:sideeffect}

\begin{figure}[t!]
\begin{adjustbox}{center}
\includegraphics[width=0.75\textwidth]{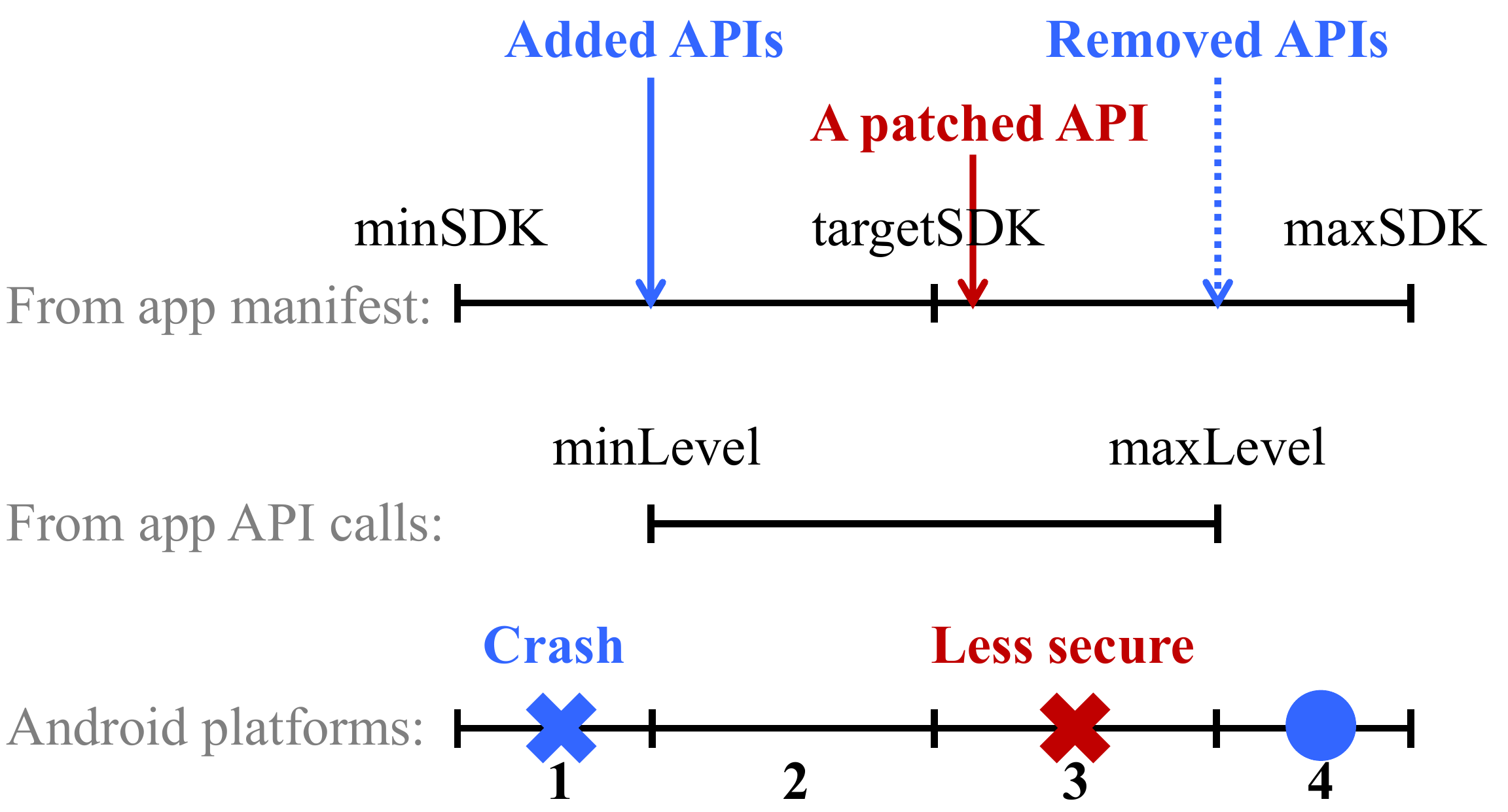}
\end{adjustbox}
\caption{Illustrating the two side effects of inappropriate \DSDK versions.}
\label{fig:demystify}
\end{figure}

\myfig~\ref{fig:demystify} illustrates two side effects of inappropriate \DSDK versions.
We first \blue{explain the used symbols}, and then describe the two side effects. 
As shown in \myfig~\ref{fig:demystify}, we can obtain $minSDK$, $targetSDK$, and $maxSDK$ from an app manifest file.
Based on the API calls of an app, we can calculate the minimum and maximum API levels it requires, i.e., $minLevel$ and $maxLevel$. 
Eventually, the app will be deployed to a range of Android platforms between $minSDK$ and $maxSDK$.

\subsubsection{Side Effect I: Causing Runtime Crashes}
\label{sec:effect1}

The blue part of \myfig~\ref{fig:demystify} shows two scenarios in which inappropriate \DSDK versions could cause compatibility-related inconsistency.
The first scenario is $minLevel > minSDK$, which means a new API is introduced after the $minSDK$.
Consequently, when an app (i) runs on Android platforms between $minSDK$ and $minLevel$ (marked as block 1 in \myfig~\ref{fig:demystify}) and (ii) executes that new API, it will crash.
We verified this case using the \texttt{VpnService} class'\red{s} \texttt{addDisallowedApplication()} API, which was introduced on Android 5.0 at API level 21.
We invoked this API in the MopEye app \cite{MopEye17} and ran \red{it} on an Android 4.4 device.
When the app executed the \texttt{addDisallowedApplication()} API, it crashed with the \path{java.lang.NoSuchMethodError} exception.

The second scenario is $maxSDK > maxLevel$, which \red{suggests that} an old API is removed at the $maxLevel$.
Although it looks \red{like} the app would crash when it runs on Android platforms between $maxLevel$ and $maxSDK$, it turns out that Google intentionally keeps the forward compatibility (by keeping those removed APIs in the framework as hidden APIs) so that developers have no concern in over-setting \maxSDK.
As a result, this scenario would not cause runtime method availability errors.
Therefore, \red{in this paper,} we measure only the first scenario of compatibility inconsistency that can cause runtime crashes.

\subsubsection{Side Effect II: Making Apps Vulnerable}
\label{sec:effect2}
The red part of \myfig~\ref{fig:demystify} shows the scenario \red{where} inappropriate \DSDK versions cause failure for the app \red{that should be} patched.
Supposing an app calls an API whose implementation is vulnerable at $targetSDK$, even when the app runs on an updated Android system (with API level $> targetSDK$), Android still \red{exhibits} the compatibility behaviors, i.e., the vulnerable implementation of the API at $targetSDK$ in this case.

\begin{table}[h!]
\caption{Vulnerable APIs or components on Android and their patched versions.}
\begin{adjustbox}{center}
\begin{tabular}{ c | c | c }

\hline
Vulnerable APIs/Components & Patched SDKs (Android) & Changed Behavior \tabularnewline
\hline
\hline

\texttt{file://} scheme in WebView  & $targetSDK \geq 16$ (4.1+) & Fix flawed same-origin policy~\cite{FileCross14} \tabularnewline
\hline

Content Provider component          & $targetSDK \geq 17$ (4.2+) & \red{Disable the default exposure}~\cite{ContentScope13} \tabularnewline  
\hline

\texttt{addJavascriptInterface()}   & $targetSDK \geq 17$ (4.2+) & Stop Java reflection for RCE~\cite{addJavaScriptInterface14} \tabularnewline     
\hline

\multirow{2}{*}{\texttt{PreferenceActivity} class} 
& \multirow{2}{*}{$targetSDK \geq 19$ (4.4+)} 
& Add \texttt{isValidFragment()} for apps \tabularnewline
&  & to prevent Fragment Hijacking~\cite{TargetFragment16} \tabularnewline 
\hline

\multirow{2}{*}{\texttt{javascript:} in WebView} 
& \multirow{2}{*}{$targetSDK \geq 19$ (4.4+)} 
& JavaScript URLs are executed in \tabularnewline
&  & a separate WebView context~\cite{MoST15} \tabularnewline 
\hline

\texttt{Context.bindService()} & $targetSDK \geq 21$ (5.0+) & Do not accept Implicit Intents~\cite{ImplicitService17} \tabularnewline     
\hline

\end{tabular}
\end{adjustbox}
\label{tab:vulnerableAPI}
\end{table}

Table~\ref{tab:vulnerableAPI} summarizes the previously reported vulnerable APIs or components on Android and their patched versions.
They were all wide-spread API-level vulnerabilities on Android, causing significant security impacts.
Although by-default fixes were subsequently provided at the API level, as shown in Table~\ref{tab:vulnerableAPI}, they often require developers' cooperation at the app level (e.g., updating app configuration).
Otherwise, vulnerabilities could still appear even on patched versions of Android~\cite{ImplicitService17}~\cite{SCLib18}.
In our context, an app could be exploited if they invoke the vulnerable APIs without declaring the updated \aimSDK.
As a result, analyzing these ``old'' vulnerabilities is still worthwhile and could demonstrate the security impact of our study.

In this paper, we specifically measure the vulnerable \texttt{addJavascriptInterface()} API for two reasons.
First, it has a clear API pattern for inconsistency measurement, while other cases in Table~\ref{tab:vulnerableAPI} involve multiple component-level factors that could cause a vulnerability.
Second, the \texttt{addJavascriptInterface()} API \red{gives rise to} the most serious security issue~\cite{LUDroid20}.
By exploiting this API, attackers are able to inject malicious code, which can cause remote code execution (e.g., stealing sensitive information from a victim app or SD card).
Google later fixed this weakness on Android 4.2 and above.
However, if an app sets the \aimSDK lower than 17 and \red{also} calls this API, the system will still render the vulnerable API behavior even when running on Android 4.2+.
Such vulnerable app examples are available at \url{https://sites.google.com/site/androidrce/}.

\section{Methodology}
\label{sec:method}

To understand how \DSDK versions are used in the wild and the pervasiveness of the two side effects in real apps, we propose an automatic approach for a systematic measurement.
In this section, we first \red{present} an overview of our methodology, and then its two \red{main} analysis phases.

\subsection{Overview}
\label{sec:overview}

\begin{figure}[t!]
\begin{adjustbox}{center}
\includegraphics[width=0.95\textwidth]{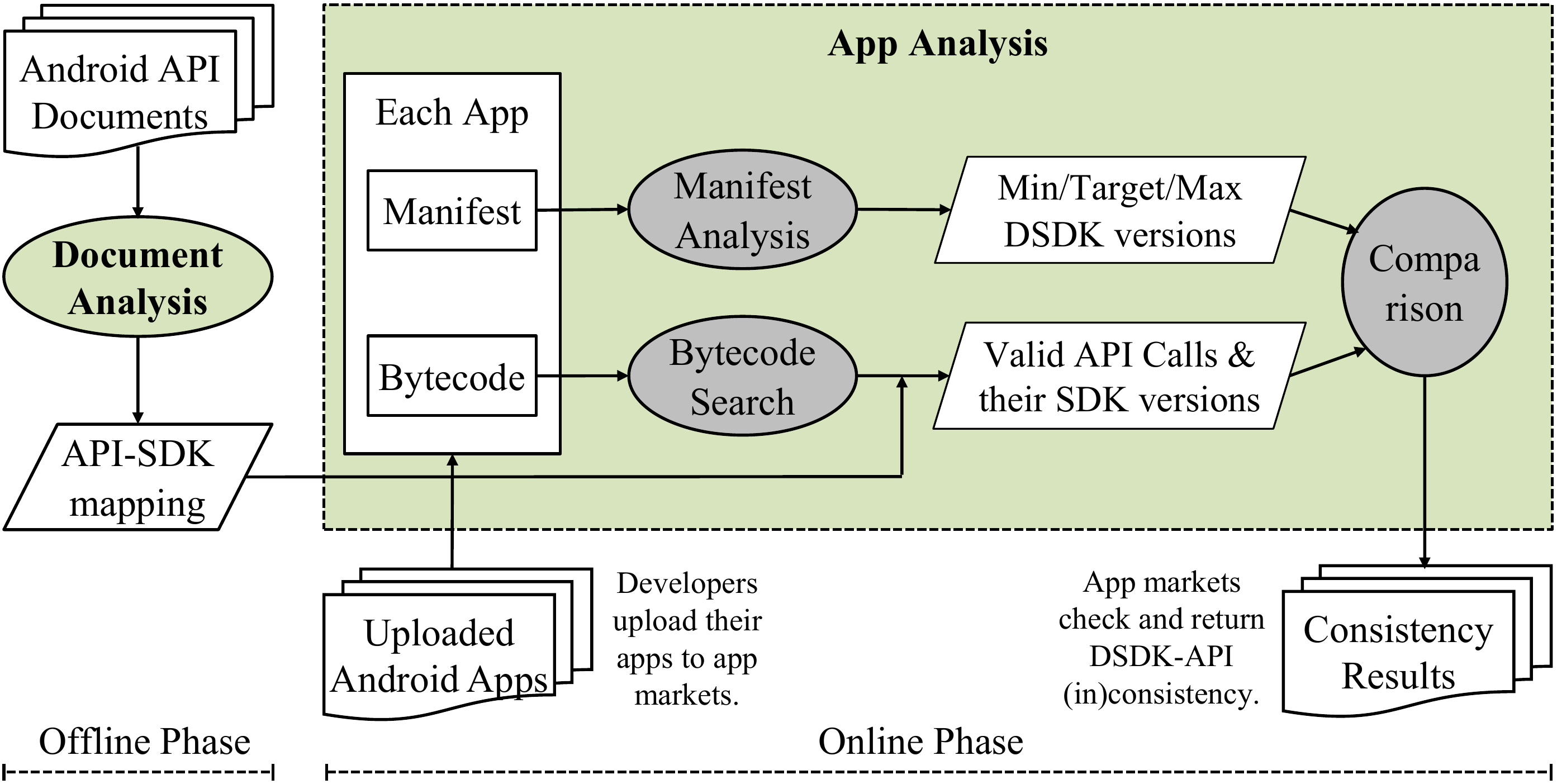}
\end{adjustbox}
\caption{The overview of our methodology.}
\label{fig:overview}
\end{figure}

\red{Our main design goal is to help the app markets} timely notify developers the \DSDK inconsistency in their apps.
\myfig~\ref{fig:overview} illustrates its overall design, where the app analysis part is conducted in the online phase.
Since our app analysis requires the \textit{API-SDK mapping} as an input (for calculating API levels of all valid API calls in an app), we further conduct Android API document analysis to build a mapping between each Android API and their corresponding SDK versions (or API levels).
As this step is performed only once, we include it in the offline phase.

The \red{majority} of our approach \blue{is designated for the online vetting of apps}.
Specifically, whenever developers upload a new or updated app to app markets, we first unzip this app to obtain its bytecode DEX file(s).
We then launch manifest analysis to robustly retrieve an app's declared SDK versions.
For bytecode analysis, \red{the} novelty is \red{that we} propose a lightweight bytecode search, instead of heavyweight dataflow analysis, to extract valid API calls.
Finally, we leverage the API-SDK mapping to calculate the range of \red{the} corresponding API levels from API calls, and compare them with the declared SDK versions.
The output is the (in)consistency results between declared SDK versions and API calls.
It is worth noting that \textit{multiple-apk} analysis~\cite{WASA17} is no longer needed in our online analysis, because app markets control all versions of APKs and multiple-apk mechanism is largely used for different hardware configuration~\cite{multiapk}.

\subsection{Offline Phase: API Document Analysis}
\label{sec:DocAnalysis}

In this subsection, we present our offline phase in detail, including both \red{the} methodology and results of API document analysis.

\begin{figure}[t!]
\begin{adjustbox}{center}
\begin{minipage}{1\textwidth}
        \centering
        \includegraphics[width=0.9\textwidth]{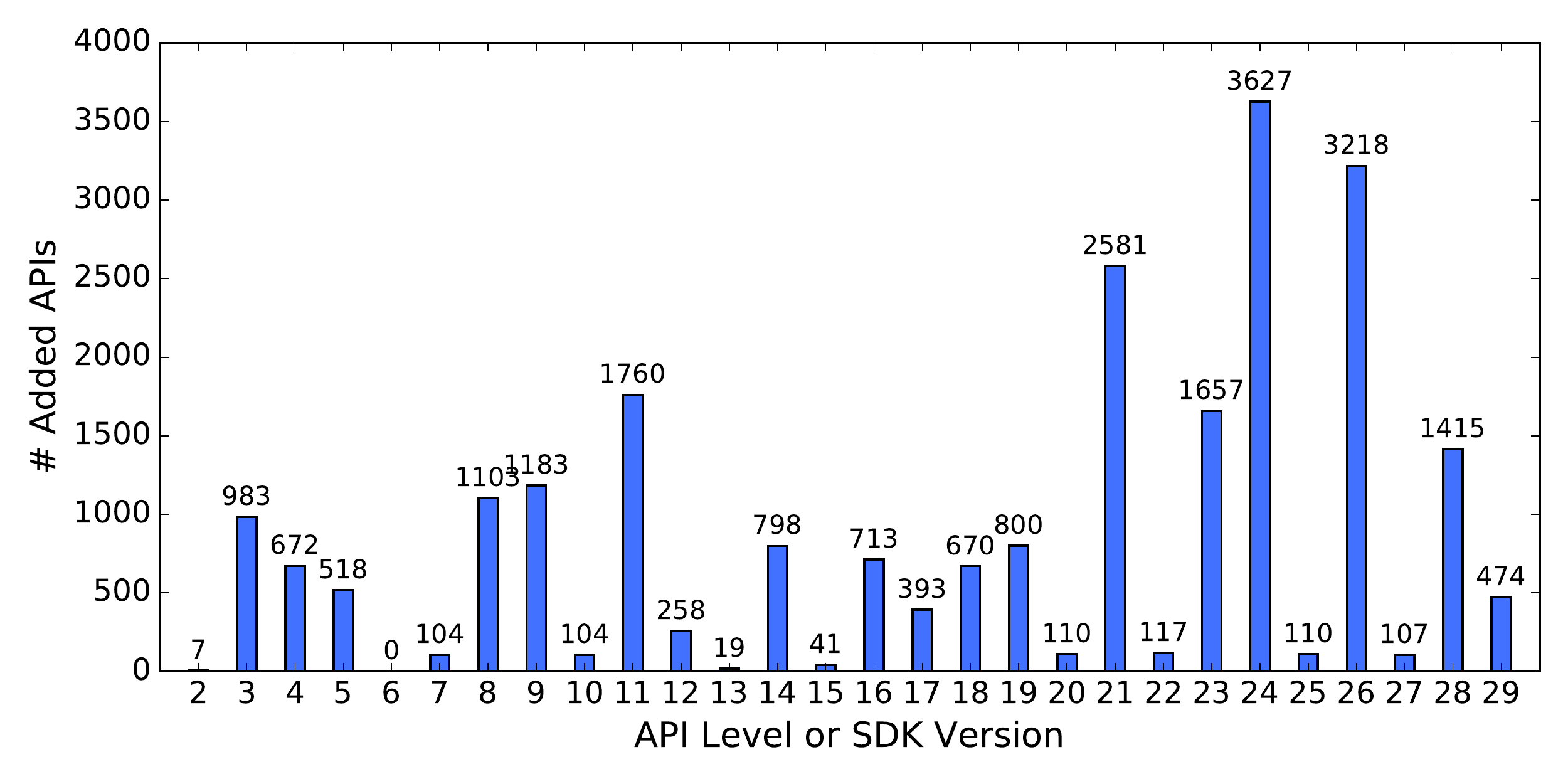}
        \caption{The distribution of added Android APIs across different SDK versions.}
        \label{fig:addedAPI}
    \vspace{3.5ex}
        \centering
        \includegraphics[width=0.9\textwidth]{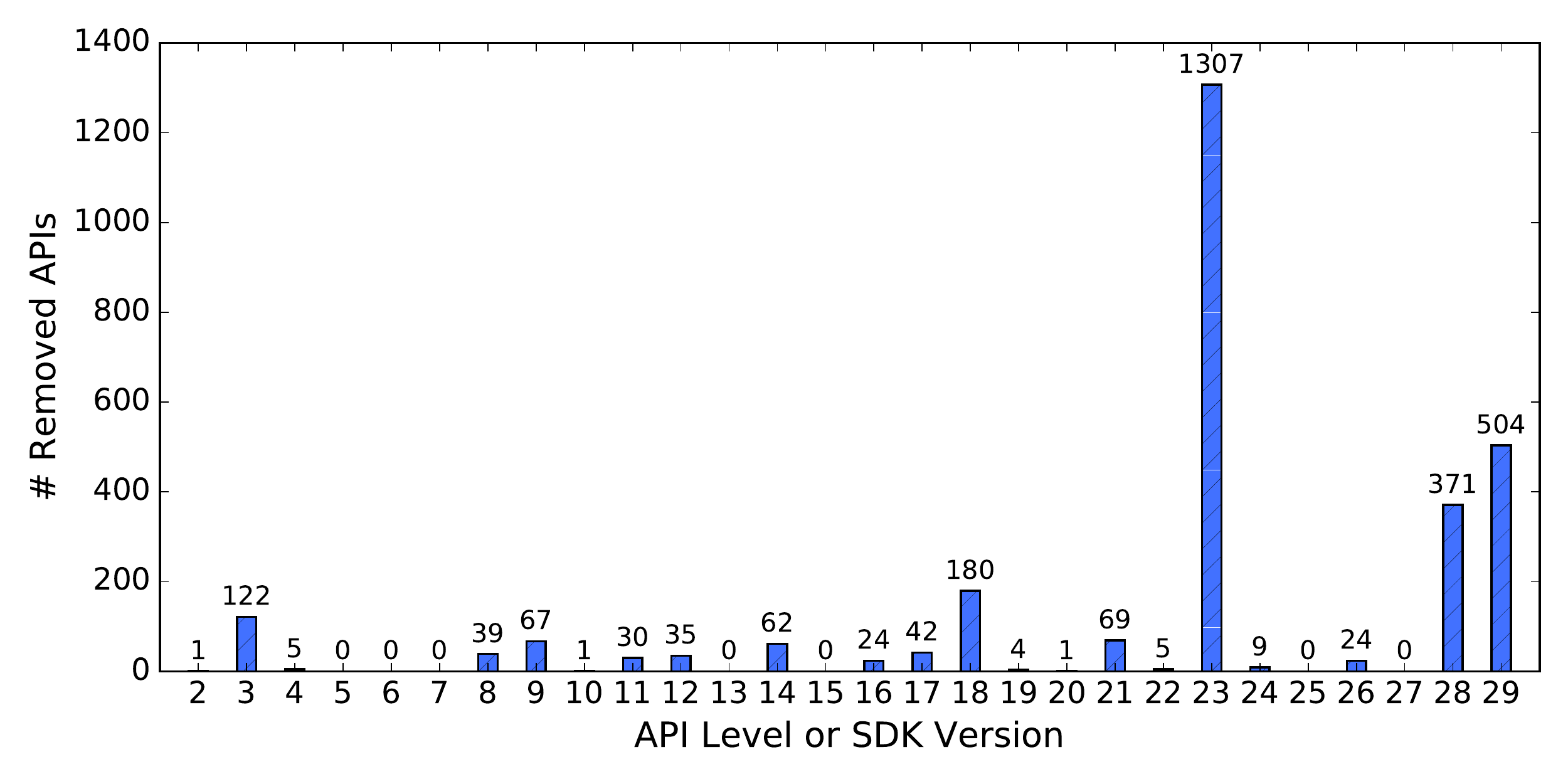}
        \caption{The distribution of removed Android APIs across different SDK versions.}
        \label{fig:removedAPI}
    \vspace{3.5ex}
        \centering
        \includegraphics[width=0.9\textwidth]{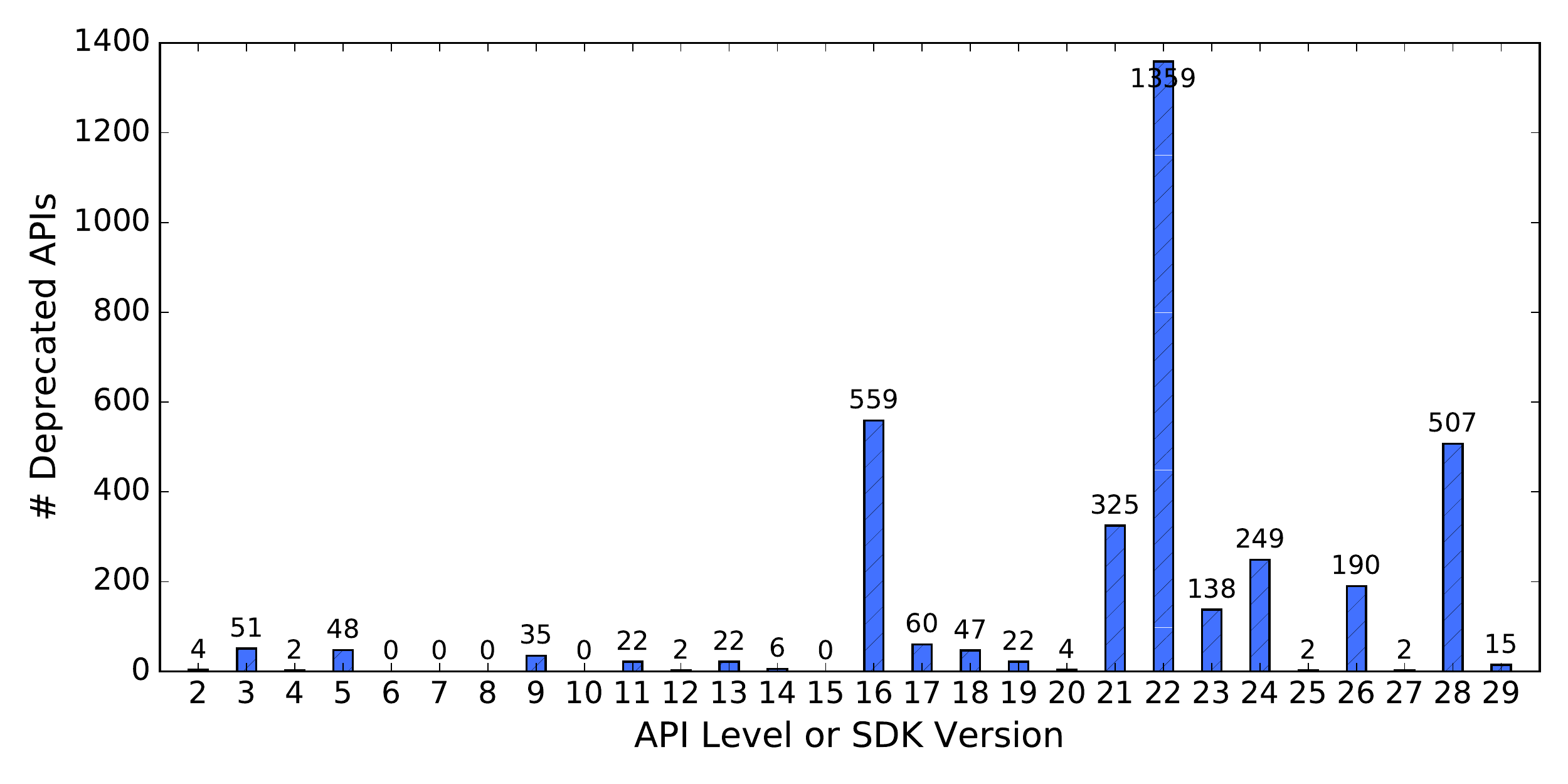}
        \caption{The distribution of deprecated Android APIs across different SDK versions.}
        \label{fig:deprecatedAPI}
\end{minipage}
\end{adjustbox}
\end{figure}

\textbf{Building the API-SDK mapping.}
There are two potential approaches for building the API-SDK mapping.
One is to analyze Android API documents by parsing a SDK document called \texttt{api-versions.xml}.
A previous API study~\cite{ICSM13} and our preliminary study~\cite{WASA17} leveraged this approach to obtain initial and added APIs, but \red{they did not cover removed and deprecated APIs because of no such information in the \texttt{api-versions.xml} file}.
\red{Hence,} they also needed to analyze the HTML files in the \texttt{api\_diff} directory, which is unfortunately error-prone~\cite{WASA17}.
The other approach is to directly retrieve the API-SDK mapping from each SDK \texttt{jar} file.
However, different SDK releases under the same API level may have some API differences, and there are over 600 releases\footnote{See tags in \url{https://android.googlesource.com/platform/frameworks/base.git/+refs}.} for \red{the} 28 API levels at the time of our writing.
As a result, conflicted API mapping\red{s} could be recorded, e.g., marking the \texttt{Gravity.getAbsoluteGravity} API removed in SDK version 16 and then added back in version 17~\cite{CiD18}.

Fortunately, we find that the first approach now \blue{covers} all kinds of Android APIs.
Specifically, the latest \texttt{api-versions.xml} file released in Android 9 SDK records all added, removed, and deprecated APIs.
Therefore, we can simply parse this file to obtain a \blue{complete} API-SDK mapping.

\textbf{Document analysis results.}
With the accurate API-SDK mapping, we are able to present \red{a} comprehensive evolution of Android APIs across different SDK versions.
\myfig~\ref{fig:addedAPI},~\ref{fig:removedAPI}, and~\ref{fig:deprecatedAPI} plot the distribution of added, removed, and deprecated Android APIs from API level~2 to the very recent API level~29, respectively.
Overall, we find that 26,466 (67.8\%) out of \red{the total} 39,034 Android APIs are changed\red{. This result} indicates that Android APIs evolve dramatically \red{during} the whole evolution.

The biggest change in \red{the} Android API evolution is to add 23,542 APIs since level~2, as shown in \myfig~\ref{fig:addedAPI}.
Specifically, Android 7.0 (API level 24) changed most, with 3,627 new APIs introduced.
Android 8.0 (API level 26) and Android 5.0 (API level 21) also introduce\red{d} a significant number of new APIs, with 3,218 and 2,581 APIs added, respectively.
Other versions of platforms with a large number of added APIs are Android 3.0 (API level 11), Android 6.0 (API level 23), and Android 9.0 (API level 28).
These new Android APIs bring a huge risk of compatibility inconsistency, causing runtime crashes on lower versions of Android.
In particular, we notice that over half (13,306, 56.5\%) of all \red{the} added APIs are introduced since Android 5.0, giving them \red{a} higher chance of causing compatibility inconsistency than the rest of \red{added} APIs.

In contrast, only 4,830 (18.2\%) APIs involve the removal change (i.e., removed or deprecated; some of them are also introduced after API level 2), with 3,671 APIs deprecated and 2,902 APIs finally removed.
According to \myfig~\ref{fig:removedAPI} and~\ref{fig:deprecatedAPI}, the biggest removal happens in Android 5.1 and 6.0 (API level 22 and 23), with 1,359 APIs deprecated and 1,307 APIs removed afterwards.
Moreover, Android 9.0 (API level 28) deprecates 507 APIs and its next version (API level 29) removes 504 of them, which suggests that Google plans to remove a large number of APIs in the release of Android 9.0.
Additionally, although Android 4.1 (API level 16) deprecated 559 APIs, only 222 APIs were removed in the subsequent Android 4.2 and 4.3.

To sum up, 23,542 (60.3\%) out of all the 39,034 Android APIs are introduced at a SDK version other than the initial Android SDK version (i.e., API level 1), which brings a high risk for developers to under-set the \minSDK attribute. 
On the other hand, much fewer Android APIs, 7.4\% of all APIs, are mapped to a range of SDK versions that have an upper limit \red{(i.e., deleted in recent SDK versions)}.

\subsection{Online Phase: Android App Analysis}
\label{sec:AppAnalysis}

In this subsection, we present three major modules in the online analysis phase, namely manifest analysis, bytecode search, and consistency comparison in \myfig~\ref{fig:overview}.

\subsubsection{Retrieving \DSDK Versions via Manifest Analysis}
\label{sec:appManifest}

To robustly retrieve \DSDK versions from all apps, we propose a new manifest analysis method that leverages \aapt (Android Asset Packaging Tool)~\cite{aapt} to retrieve \DSDK \textit{directly} from each app without extracting and decoding the manifest file.
This method is more robust than the traditional \apktool-based manifest extraction~\cite{apktool}\red{, which} requires to extract and decode the manifest into a plaintext file. 
Indeed, our \aapt-based approach can successfully analyze all 22,687 apps, whereas a previous work~\cite{ECVDetector14} showed that \apktool failed six times in the analysis of just 1K apps.
Specifically, we utilize the \texttt{dump baging} command in \aapt to extract \red{the} \DSDK versions.
\blue{In this way, we can directly retrieve the correct \texttt{DSDK} versions without analyzing raw manifest files. Therefore, even \red{when} an app contains old or unreferenced manifest files, it would not affect our analysis.}


In the course of implementation and evaluation, we observed and handled two kinds of special cases.
First, some apps define \minSDK multiple times, for which we only extract the first value.
Second, we apply the default rules (see \mysec\ref{sec:declared}) for apps without \minSDK and \aimSDK defined.
More specifically, we set the value of \minSDK to 1 if it is not defined, and set the value of \aimSDK (if it is not defined) using the \minSDK value.

Besides retrieving \DSDK, our manifest analysis also parses all components registered in the manifest to generate a list of valid components and their root (Java) class names.
\red{This} information will be used in the app analysis module to exclude uninvoked third-party libraries.
Specifically, we execute the \texttt{dump xmltree} command in \aapt to output all component information.
In the process of parsing these components, we also generate their root class names according to this rule:
if the component class does not overlap with the app package or \texttt{<application>} name (i.e., this class could be from a third-party library), we record the entire class name as the root class; otherwise, only the leading two or three name portions are treated as the root class.

\subsubsection{Extracting Valid API Calls via Bytecode Search}
\label{sec:appAPIcall}

The main module in our app analysis is to extract valid API calls.
A valid API call \red{is a call \textit{not}} guarded by the \texttt{VERSION.SDK\_INT} checking (a mechanism developers can use to invoke an API only in certain Android platforms).
It \red{should also not appear} in uninvoked third-party libraries that are essentially dead code. 
To guarantee the scalability for online vetting, we propose a lightweight \textit{bytecode search}, instead of dataflow-based approaches, for app analysis, because existing Android dataflow analyses, notably FlowDroid~\cite{FlowDroid14} and Amandroid~\cite{Amandroid14}, are expensive even when analyzing medium-size\red{d} apps, e.g., requiring $\sim$4 minutes for just an app of size 8MB~\cite{IctApiFinder18}.

Moreover, we operate on the original Android bytecode level without decompiling app bytecodes\red{, which helps} \red{reduce} false negatives.
This is because the process of transforming or decompiling Android app bytecode into an intermediate representation (usually Java bytecode) is not fully accurate~\cite{RetargetDex12}.
As a result, many previous studies~\cite{AppContext15}~\cite{MudFlow15}~\cite{MaMaDroid17}~\cite{HSOMiner17} often failed to handle some apps, causing false negatives in their analysis.
In contrast, by directly analyzing app bytecodes, we robustly process all 22,687 popular apps in our dataset.
Specifically, we leverage the \dexdump tool~\cite{OpenPort19} to translate compressed bytecodes into plain bytecode texts (similar to using \texttt{objdump} to generate assembly code texts), upon which we can then launch bytecode search to extract valid API calls.   
Note that \dexdump, as an official Android SDK tool, is very robust, and it does not generate intermediate representation.
We also dump (multiple) app bytecodes into a (combined) plaintext~\cite{OpenPort19} to handle multidex~\cite{multidex}, a special bytecode format often skipped by prior works but indeed common in modern apps --- 5,008 apps in our dataset split their bytecodes into multiple files.
Hence, we avoid another common source of false negatives.

In the rest of this subsection, we first introduce the basic bytecode search mechanism before describing our bytecode search of \texttt{VERSION.SDK\_INT} checking and vulnerable API calls in details.
We then explain how we exclude uninvoked third-party libraries during the search process.

\begin{figure}[t!]
\begin{adjustbox}{center}
\includegraphics[width=0.3\textwidth]{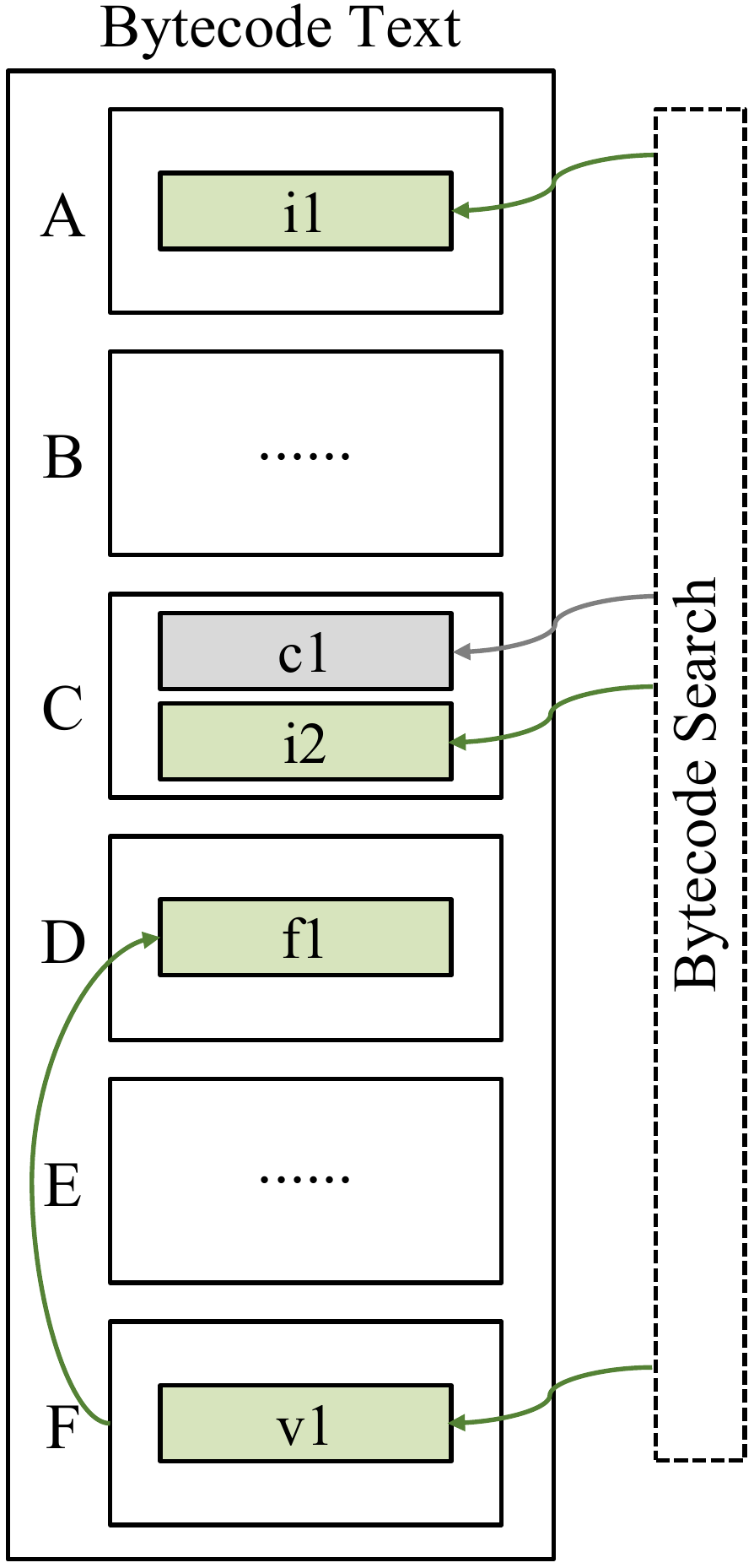}
\end{adjustbox}
\caption{A high-level overview of our bytecode search mechanism.}
\label{fig:bytecodeSearch}
\end{figure}

\textbf{The basic bytecode search mechanism.}
\myfig~\ref{fig:bytecodeSearch} shows a high-level overview of our bytecode search mechanism.
The bytecode text outputted by \dexdump is a sequence of code statements, hierarchically organized by different class and method bodies.
In \myfig~\ref{fig:bytecodeSearch}, we show six method bodies (from method A to method F), where their corresponding class bodies are omitted for simplicity.
As illustrated in the figure, our bytecode search scans these methods to locate inconsistent API calls (e.g., call site i1 and i2 in method A and C, respectively) and vulnerable API calls (e.g., call site v1 in method F).
We can perform further search to determine in which class an interested method is invoked, e.g., \myfig~\ref{fig:bytecodeSearch} shows that method F (containing vulnerable API call v1) is called by another method D.
Besides \red{the} method search, we can also launch \texttt{if} statement search to locate conditional checking, e.g., statement~c1 that surrounds call site i2 in method~C.

\textbf{Searching \texttt{VERSION.SDK\_INT} checking.}
As mentioned earlier in this subsection, developers can use \texttt{if} statements with \texttt{VERSION.SDK\_INT} checking to invoke an API only in certain Android platforms, thus avoid\red{ing} the inconsistency problem.
Listing~\ref{lst:sdkChecking} shows an example of \texttt{VERSION.SDK\_INT} checking, which invokes the \texttt{addDisallowedApplication()} API (introduced since API level 21) only on Android 5.0 and above.
To avoid such false positives, our approach must \red{handle the} \texttt{VERSION.SDK\_INT} checking.

\begin{lstlisting}[
float=t!,
%language=XML,
basicstyle=\ttfamily,
style=customJava,
label={lst:sdkChecking},
caption={An example of \texttt{VERSION.SDK\_INT} checking.}
]
VpnService.Builder builder = new VpnService.Builder();
if (VERSION.SDK_INT >= VERSION_CODES.LOLLIPOP) {
    builder.addDisallowedApplication(Constant.PkgName);
}
\end{lstlisting}

Our strategy is to \red{perform} both API call and \texttt{VERSION.SDK\_INT} checking search and see whether the two search results overlap in the same method.
For example, in \myfig~\ref{fig:bytecodeSearch}, our bytecode search locates both checking statement c1 and API call i2 in method C.
Since these two search results overlap and API call i2 is invoked below checking statement c1, we are thus confident that this API call has been guarded with a corresponding \texttt{VERSION.SDK\_INT} checking.
\blue{Moreover, according to a recent study~\cite{IctApiFinder18}, 88.65\% of the \DSDK checking usages directly compare the \texttt{VERSION.SDK\_INT} variable with a constant Android version number, which makes our bytecode search strategy appropriate.}

\textbf{Searching vulnerable API calls.}
For a vulnerable API call, we further employ bytecode search to determine whether it is initialized by app's own code or library code.
This is particularly important for the vulnerable API studied in this paper, namely \texttt{addJavascriptInterface()}, because a previous study has shown that over 47\% of top 40 ad libraries create their Javascript Interfaces~\cite{addJavaScriptInterface14}.  
Specifically, after locating vulnerable API call v1 in method F, we further search the invocation(s) of method F to check its origin class.

\textbf{Excluding uninvoked third-party libraries.}
An important issue during our bytecode search is to exclude uninvoked third-party libraries.
\blue{To tackle this problem, we cannot simply employ library detection (e.g., LibScout~\cite{LibScout16} and LibD~\cite{LibD17}) to exclude \textit{all} libraries, because this approach also ignores those invoked and thus valid library code.}
\red{To keep valid libraries as much as possible while minimizing the false positives raised by uninvoked libraries, we propose a lightweight yet practical approach that combines both heuristics-based component analysis and API-based bytecode search.
  Specifically, we first conservatively exclude all the code that has no relationship with app component information even though some of them might be functionality-supporting code.
We achieve this by performing manifest analysis and generating root classes for all registered components, as mentioned in \mysec\ref{sec:appManifest}.
A class whose code does not appear in any root class is thus recognized as an uninvoked library or dead code.
Note that even for a valid third-party library, only its registered components will be analyzed because not all code in a library will be invoked by the main app.}
\blue{Furthermore, when a candidate API call is going to be reported during the detection phase, we launch one more bytecode search to double check its invocations.}
\red{Eventually, the identified inconsistency cases will be confirmed by developers, and as we will discuss in \mysec\ref{sec:discuss}, the effort of performing such checking is minimal}.
\red{In this way, we consider valid third-party libraries but also minimize their potential false positives, without relying on the expensive dataflow-based analysis that does not meet the objective of online vetting in app markets.}

\subsubsection{Calculating API Levels and Comparing Their Consistency with DSDKs}
\label{sec:calculateANDcompare}

With the extracted API calls, we use the API-SDK mapping to compute the range of corresponding API levels (i.e., from \minLevel to \maxLevel, as explained in \mysec\ref{sec:sideeffect}).
The \minLevel of an app is the maximum of all its valid API calls' corresponding \minLevel values (i.e., all correspondingly added SDK versions), while the \maxLevel of an app is the minimum of all valid API calls' corresponding \maxLevel values (i.e., all correspondingly removed SDK versions).
If an API is never removed, we set a large flag value (e.g., 100,000) to represent its \maxLevel value.

We then compare \red{the} extracted \DSDK values with the calculated API levels to obtain the following two kinds of inconsistency (as previously mentioned in \mysec\ref{sec:sideeffect}):
\begin{itemize}
\item $\minSDK < \minLevel$: the \minSDK is set too low and the app would crash when it runs on platform versions between \minSDK and \minLevel.

\item $\aimSDK < \maxLevel$: the \aimSDK is set too low and the app could be updated to the version of \maxLevel. If the \maxLevel is infinite, the \aimSDK could be adjusted to the latest Android version.

\end{itemize}

\section{Evaluation}
\label{sec:evaluate}

Our evaluation aims to answer the following \blue{five} research questions:
\begin{description}
  \item[\textbf{RQ1}] What are the \textit{current \DSDK characteristics} in popular real-world apps?

  \item[\textbf{RQ2}] How pervasive is the \textit{compatibility-related inconsistency} in real-world apps?

  \item[\textbf{RQ3}] How pervasive is the \textit{security-related inconsistency} is in real-world apps?

  \item[\textbf{RQ4}] How \textit{scalable} is our inconsistency detection approach?

  \item[\textbf{\blue{RQ5}}] \blue{What is the \textit{updatablity} of the buggy apps? Are they still being maintained?}
\end{description}

\noindent
We choose popular real-world apps\red{,} instead of randomly selected apps or open-source apps, for evaluation, because they are most likely installed by regular users (according to Google Play installs).
Hence, the obtained measurement results can reflect the \DSDK practice in the wild.
In this section, we first describe how we collect such a large dataset in \mysec\ref{sec:dataset}.
Based on this dataset, we then answer \red{the} \blue{five} research questions from \mysec\ref{sec:rqDSDK} to \mysec\ref{sec:rqUPDATE}.

\subsection{Dataset}
\label{sec:dataset}

We collect popular apps on Google Play via the AndroZoo repository~\cite{AndroZoo16}, which contains a total of 3,699,731 unique\footnote{An app is unique if its package name, instead of SHA1/256, is different from other apps.} Google Play apps at the time of our crawling on 11 November 2018.
However, AndroZoo does not provide the app install information, which is \red{needed} to determine the popularity of each app.
To quickly locate popular apps, we leverage the top app lists available \red{at} \url{https://www.androidrank.org}.
Specifically, we crawled top 1,000 app in each Google Play category (49 categories in total, including 17 different game sub-categories), and recorded the package names of apps \textit{with over one million installs}. 
This allows us to obtain a list of 25,144 popular apps, 22,687 (the rest are either paid apps or not indexed by AndroZoo) of which are available on AndroZoo.
We \red{then} downloaded these 22,687 apps as our dataset.

\begin{figure}[t!]
  \subfigure[32 non-game app categories.] {
    \includegraphics[width=1.0\textwidth]{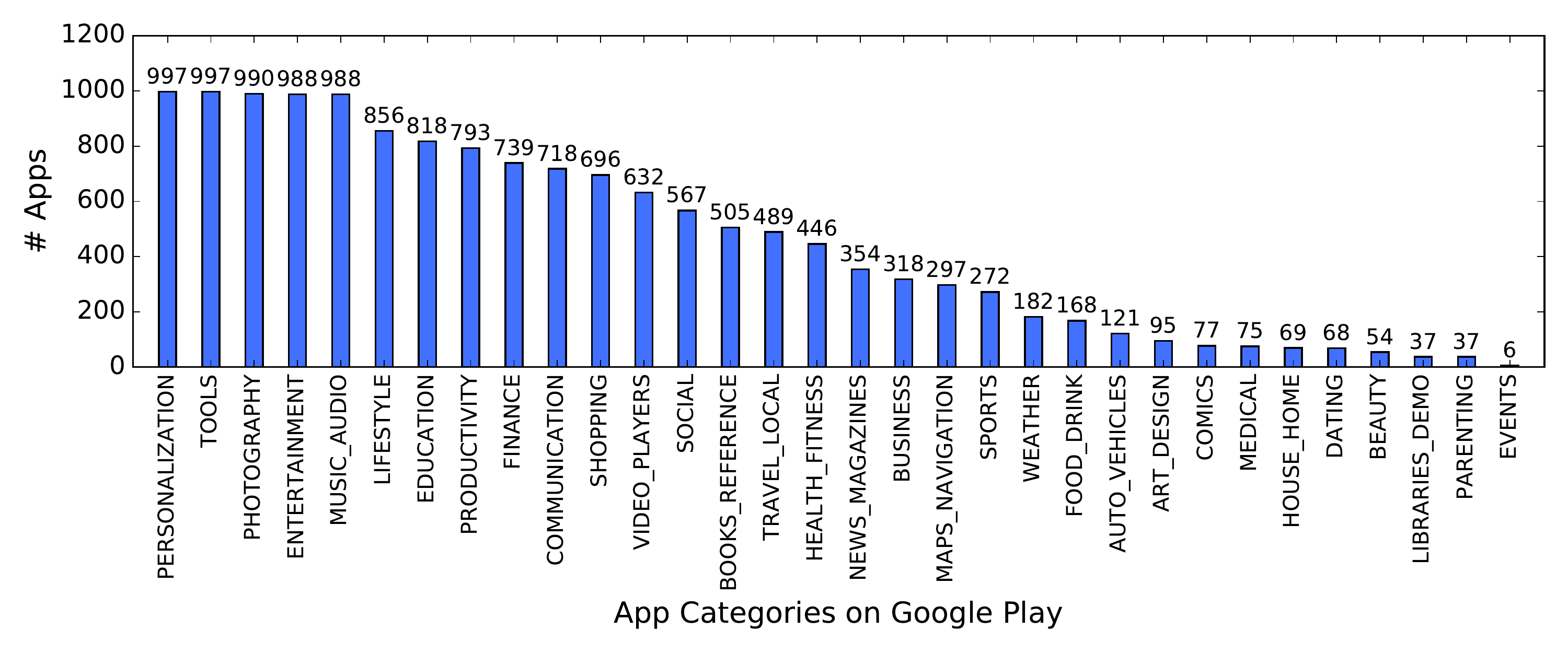}
    \label{fig:appcategoryNormal}
  }
  \subfigure[\blue{17 game app categories.}] {
    \includegraphics[width=1.0\textwidth]{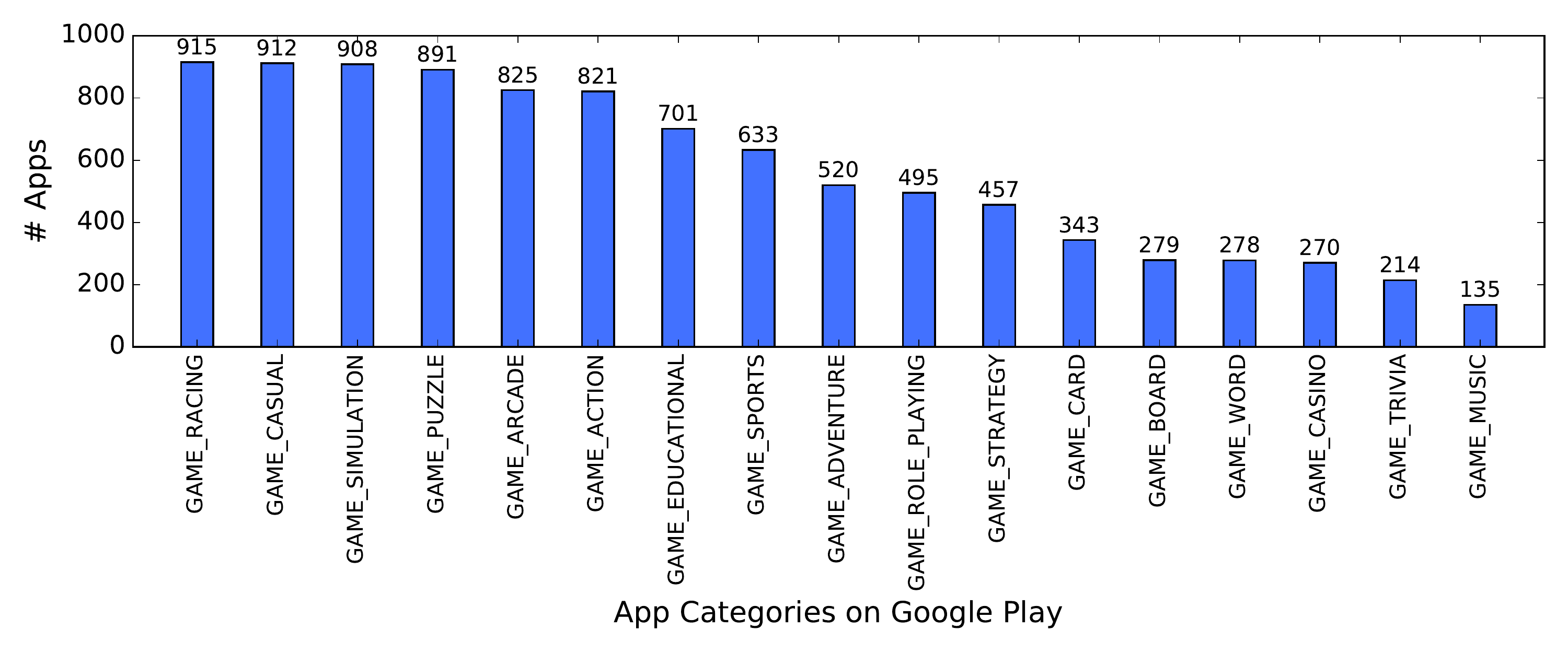}
    \label{fig:appcategoryGame}
  }
\caption{Bar charts of the distribution of popular apps across different categories.}
\label{fig:appcategoryBar}
\end{figure}

To understand \red{the distribution of} these popular apps across different app categories, we plot bar charts in \myfig~\ref{fig:appcategoryBar} that cover \blue{both} 32 non-game app categories \blue{and 17 game sub-categories}.
In particular, 17 game sub-categories contribute to a total of 10,695 popular apps, which indicates that game apps are commonly installed by real-world Android users.
According to \myfig~\ref{fig:appcategoryBar}, app categories like ``Personalization'', ``Tools'', ``Photography'', ``Entertainment'', and ``Music'' also produce a large number of popular apps, almost 1K popular apps per category. 
We notice that daily-used categories, such as ``Communication'' and ``Social'', however, do not generate an equivalent number of popular apps, with only 600 to 700 popular apps.
This is because in these categories, several very popular apps, e.g., WeChat and Facebook, dominate a large portion of \red{the} market share.
Lastly, it is also reasonable for some unpopular categories, such as ``Medical'' and ``Libraries \& Demo'', to \red{have a} limited number of popular apps.

\begin{figure}[t!]
\begin{adjustbox}{center}
\includegraphics[width=0.6\textwidth]{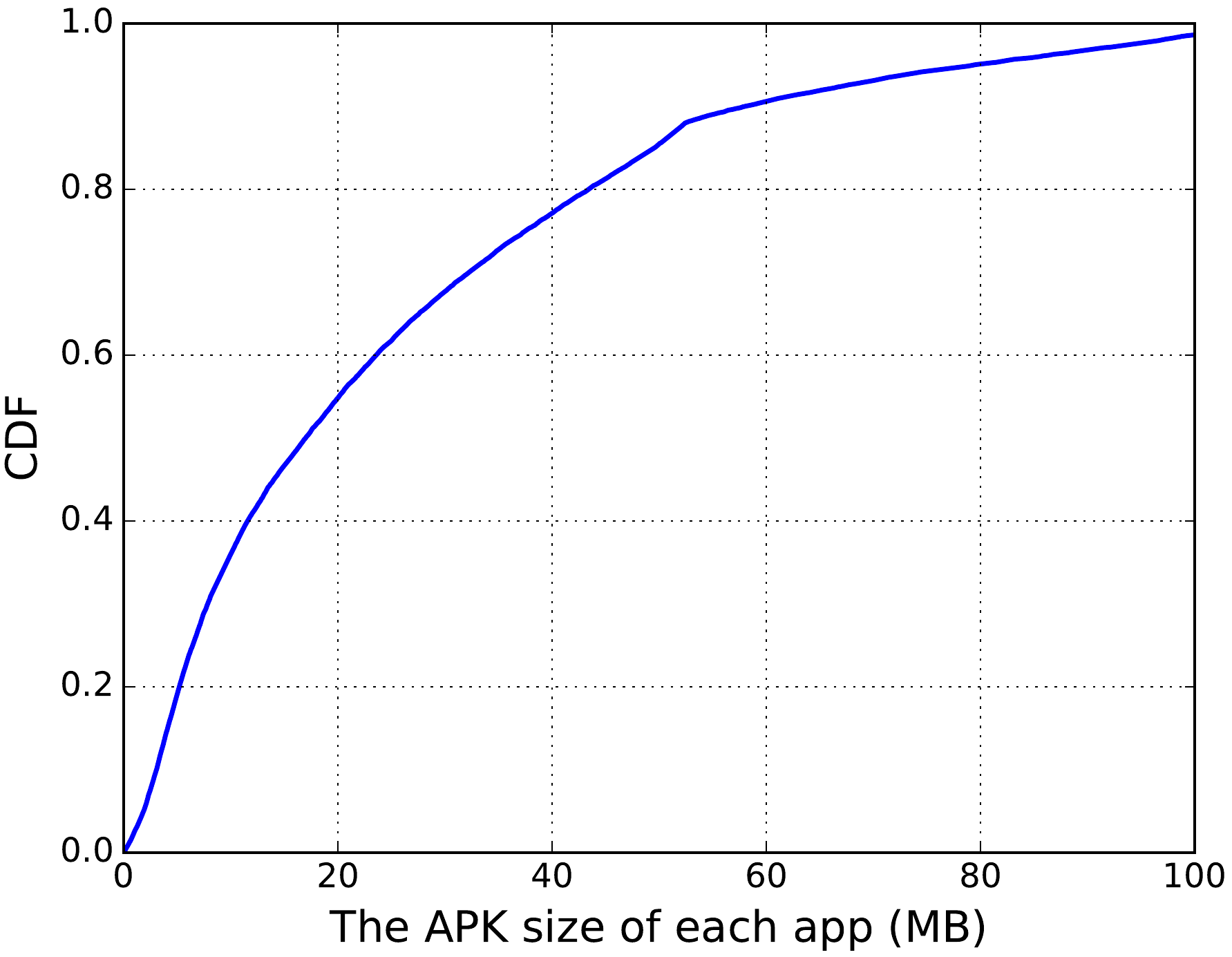}
\end{adjustbox}
\caption{CDF plot of the APK file size of each app in our dataset.}
\label{fig:APKsizeMB}
\end{figure}

It is also important to measure the distribution of app size in our dataset.
\myfig~\ref{fig:APKsizeMB} plots the CDF (cumulative distribution function) of the APK file size of each app in our dataset.
We can see that over 40\% apps have a size larger than 20MB, and over 20\% apps are even larger than 40MB each.
This indicates that a significant portion of modern apps are no longer small.
Indeed, the average app size in our dataset is 25MB, much larger than the size of apps used in several prior dataflow analysis studies (e.g., apps below 5MB were evaluated in AppContext~\cite{AppContext15}, and the maximum app size in IctApiFinder~\cite{IctApiFinder18} is 8MB).
Therefore, scalability is a key design objective for our approach, and we will evaluate it extensively in \mysec\ref{sec:rqPERFORMANCE}.

\subsection{RQ1: Characteristics of Declared SDK Versions in the Wild}
\label{sec:rqDSDK}

In this section, we report a total of four findings regarding RQ1.
We also compare these new findings with our previous results in~\cite{WASA17}, which measured a dataset of 22.7K apps crawled in 2015.

\textbf{Finding 1-1:}
\textit{Nearly all apps define the \minSDK attribute, but 4.76\% apps \red{still do not claim} the \aimSDK attribute, although this percentage has significantly dropped compared to our prior analysis in 2015.}
Table \ref{tab:nondefined} shows the number and percentage of non-defined \DSDK attributes in our dataset.
We can see that nearly all apps have defined the \minSDK attribute while \red{almost zero app} define\red{s} the \maxSDK attribute.
This result is good because, as we described in Section \ref{sec:declared}, defining \minSDK is necessary while \maxSDK is \textit{not}.
However, we also notice that 1,079 (4.76\%) apps \red{still do not claim} the \aimSDK attribute, which causes their \aimSDK values be set to the corresponding \minSDK values by default.

Fortunately, the percentage of non-defined \aimSDK has dropped \red{significantly as} compared to our prior analysis in 2015, from 16.54\% to 4.76\%.
One important factor is the popularity of Android Studio in recent years, which has become the de-facto IDE (integrated development environment) for Android app development.
Since Android Studio by default sets and enforces the \minSDK and \aimSDK attributes, the percentage of non-defined \aimSDK naturally drops and we expect that this percentage \red{will} further decrease with more apps getting updated. 

\begin{table}[t!]
\caption{The number and percentage of non-defined \DSDK attributes in our dataset.}
\begin{adjustbox}{center}
\begin{tabular}{  c| c | c}

\hline
 & \# Non-defined & \% Non-defined \tabularnewline
\hline
\hline


\minSDK    & 5          & 0.02\% \tabularnewline
\hline
\aimSDK    & 1,079      & 4.76\% \tabularnewline
\hline
\maxSDK    & 22,623     & 99.72\% \tabularnewline
\hline

\end{tabular}
\end{adjustbox}
\label{tab:nondefined}
\end{table}

\textbf{Finding 1-2:}
\textit{Some \aimSDK attributes are set to outlier values.}
We find that a total of 45 apps in our dataset declare their \aimSDK attributes as outlier values, \red{which is} close to our prior analysis \red{result} in 2015 when we encountered 55 such cases.
There are two \red{types} of outlier values.
The first is that \aimSDK is set to an API level not in the range of released SDKs.
At the time of our analysis, the valid API levels are from 1 to 28 (Android 9.0).
However, 12 apps set their \aimSDK to larger than 28, namely 29, 30, and 31.
In our prior analysis~\cite{WASA17}, one app \red{even set} its \aimSDK value to 10000.
\red{This is} probably because their developers want to always target at the latest Android SDK.

The other \red{type} of outliers is that the \aimSDK value is set to a value lower than the \minSDK value.
Normally, \aimSDK should be greater than or equal to \minSDK, but 33 apps have negative $\aimSDK$ $-$ $\minSDK$ values.
This number is almost the same as that in our prior analysis in 2015 (34 apps at that time).
In particular, one app (\texttt{com.leftover.CoinDozer}) defines its \aimSDK as 0, although its \minSDK value is 8.
We believe that this class of outliers is due to developers' mistakes in declaring \red{the} \DSDK attributes.

\begin{figure}[t!]
\begin{adjustbox}{center}
\includegraphics[width=0.9\textwidth]{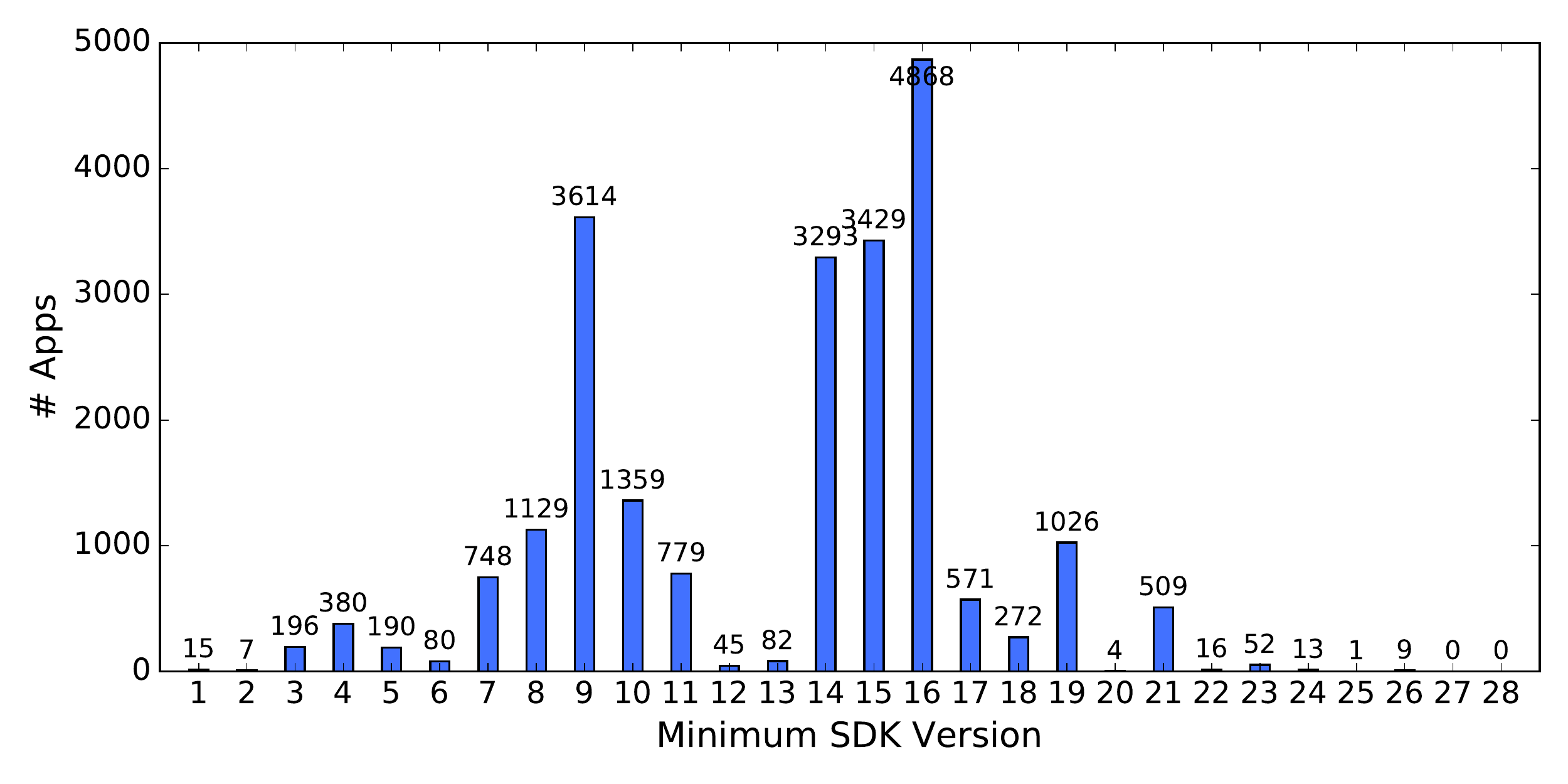}
\end{adjustbox}
\caption{Distribution of \minSDK.}
\label{fig:minSDKdistribute}
\end{figure}

\textbf{Finding 1-3:}
\textit{The minimal platform version most apps support is Android 4.1, whereas the most targeted platform version is Android 8.0. This has dramatically evolved since our last analysis in 2015.}
We first study the distribution of \minSDK.
According to \myfig~\ref{fig:minSDKdistribute}, \red{the} majority (89\%) of apps have \minSDK lower than or equal to level 16 (Android 4.1), which means that they can run on nearly all (99.5\%) Android devices in the market nowadays~\cite{dashboards}.
Specifically, the minimal platform version most apps support is Android 4.1 (level 16), while that in our last analysis in 2015 was only Android 2.3 (level 9).
However, Android 2.3 still ranks in the second place, with 3,614 apps' \minSDK targeted at.
Besides Android 4.1 and 2.3, two Android 4.0.x (level 14 and 15) platform versions are also commonly defined as apps' \minSDK.

\begin{figure}[t!]
\begin{adjustbox}{center}
\includegraphics[width=0.9\textwidth]{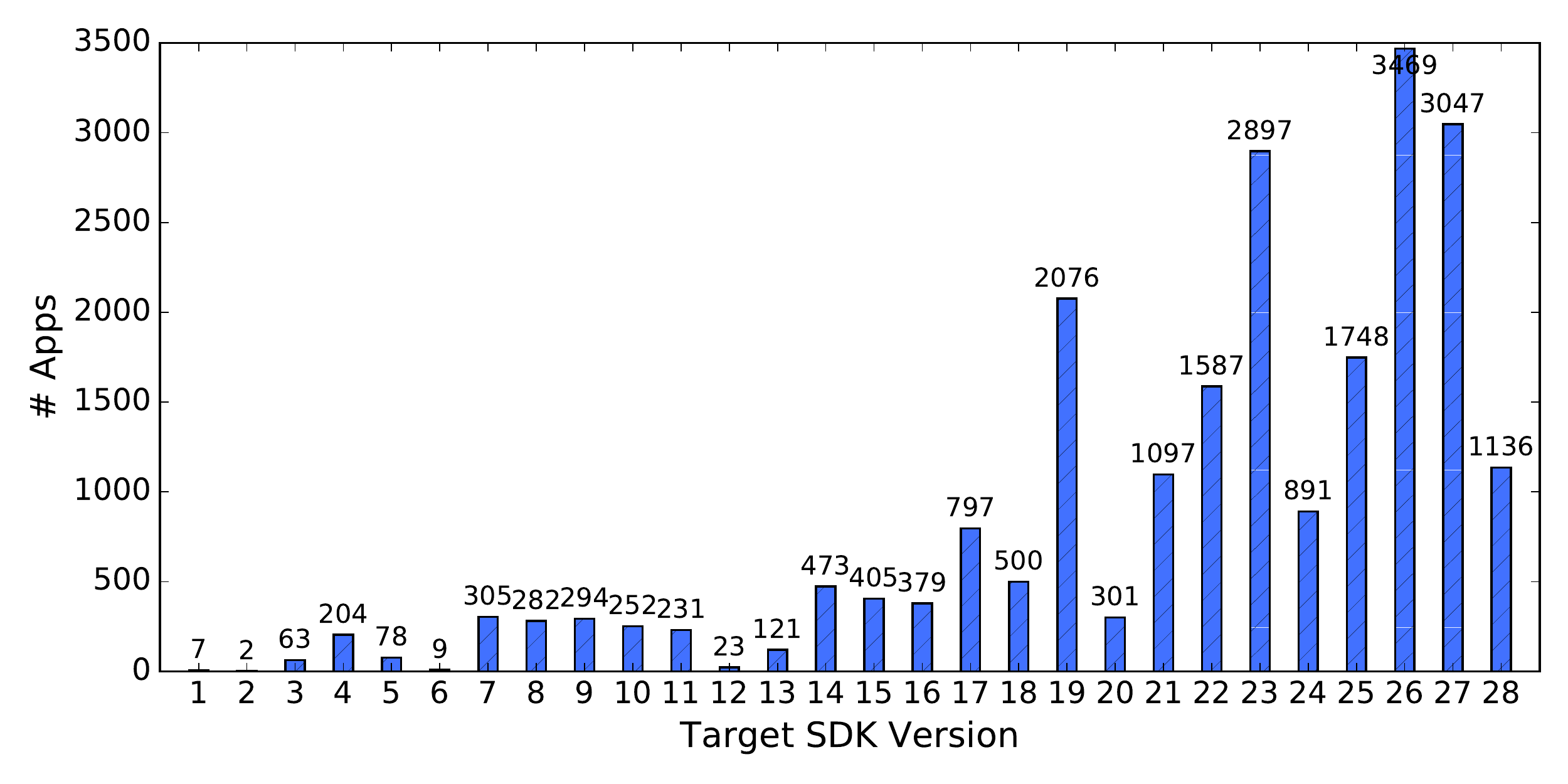}
\end{adjustbox}
\caption{Distribution of \aimSDK.}
\label{fig:tarSDKdistribute}
\end{figure}

On the other hand, \myfig~\ref{fig:tarSDKdistribute} plots the distribution of \aimSDK.
We can see that 80\% apps set their \aimSDK values \red{to} larger than or equal to level 19 (Android 4.4).
In particular, the two most targeted platform versions are the most recent Android 8.0 (level 26) and 8.1 (level 27), while those in our last analysis in 2015 were Android 4.4 and 5.0.
This suggests that modern apps keep pace with the evolution of \red{the} Android operating system.
Besides Android 8, Android 6.0 (level 23) and 4.4 (level 19) still hold a significant portion of apps with the corresponding \aimSDK setting.
Moreover, Android 7.0.x (level 24 and 25) and Android 5.0.x (level 21 and 22) also attract considerable apps being targeted at.

\begin{figure}[t!]
\begin{adjustbox}{center}
\includegraphics[width=0.6\textwidth]{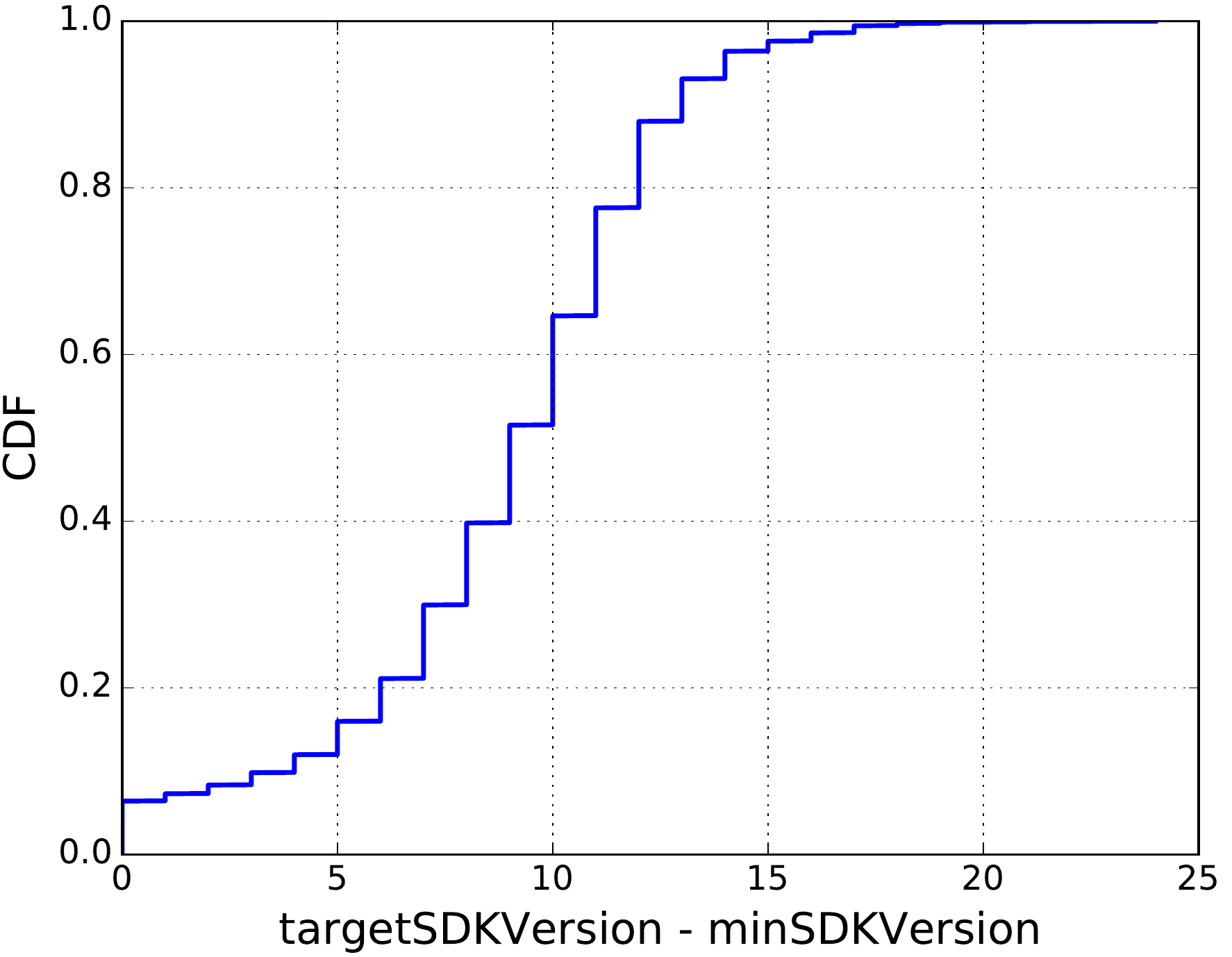}
\end{adjustbox}
\vspace{-4ex}
\caption{CDF plot of \lagSDK.}
\label{fig:lagSDK}
\end{figure}

\textbf{Finding 1-4:}
\textit{The median version difference between \aimSDK and \minSDK is 9, while that \red{in} our last analysis was 8. This 11\% increase indicates that Android apps nowadays need to support more Android platforms.}
We define a new metric called \lagSDK to measure the version difference between \aimSDK and \minSDK, as shown in Equation \ref{equ:lagversion}.
\begin{equation}
  \lagSDK = \aimSDK - \minSDK
\label{equ:lagversion}
\end{equation}
After removing \red{the} negative \lagSDK values (i.e., outliers mentioned in Finding 1-2), we draw the CDF plot of \lagSDK in \myfig \ref{fig:lagSDK}.
We first find that the median value of \lagSDK in our dataset is 9, while that \red{in} our last analysis in 2015 was 8.
It indicates that Android apps nowadays need to support more Android platforms.
This conclusion \red{is} further supported \red{by} the percentage of apps that have a \lagSDK value greater than 12.
Compared to our prior analysis, this percentage has increased from 5\% to 20\%, which clearly shows that more and more apps nowadays support a wide range of Android platforms.
On the other hand, the percentage of apps that have \red{the same value for} \aimSDK and \minSDK has also dropped from 20\% in 2015 to 6.4\% in 2018.

\subsection{RQ2: Inconsistency Results with Compatibility Effect}
\label{sec:rqCRASH}

In this section, we report three important findings regarding RQ2.
Besides presenting compatibility results as the major finding, we summarize the reduced false positives by our bytecode search as compared to the prior conference version, and show \red{that} the newly added API classes are common sources of compatibility inconsistency.

\textbf{Finding 2-1:} 
\textit{\red{Around 35\%} apps under-set the \minSDK value, causing them \red{potentially} crash when running on lower versions of Android platforms. Fortunately, only 11.3\% apps could crash on Android 6.0 and above.}
As explained in \mysec\ref{sec:calculateANDcompare}, the compatibility inconsistency happens if \minSDK is less than \minLevel.
In our experiments, we \red{thus} count the number of API calls that have higher API level than \minSDK in each app, and denote it by \minOverNum.
The higher value an app's \minOverNum is, the more likely that this app has the compatibility inconsistency.

\begin{figure}[t!]
\begin{adjustbox}{center}
\includegraphics[width=0.6\textwidth]{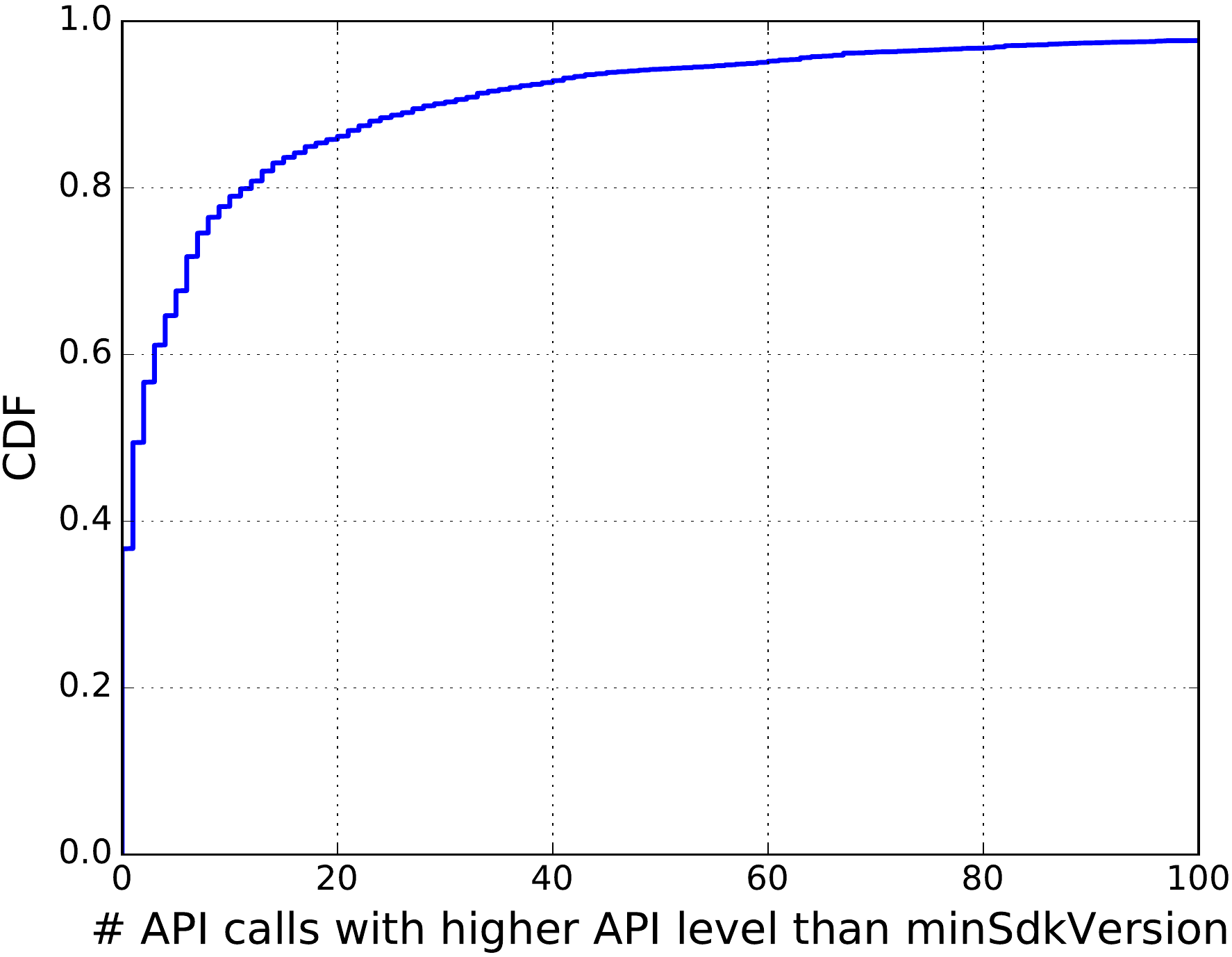}
\end{adjustbox}
\caption{CDF plot of \minOverNum in each app.}
\label{fig:minOverNum}
\end{figure}

\myfig~\ref{fig:minOverNum} shows the CDF plot of \minOverNum in each app.
We find that 14,363 (63.3\%) apps have at least one API call that has higher API level than the corresponding \minSDK.
To further increase the confidence of our analysis, we count \red{the} 8,019 (35.4\%) apps \red{that} invoke over five different API calls with higher API levels than corresponding \minSDK.
Therefore, we estimate that \red{around 35\%} apps could crash when running on lower versions of Android platforms because they under-set the \minSDK value.
Fortunately, we find that the number of inconsistency warnings per app reported by our bytecode search is well manageable for developers --- 77.8\% of the 14,363 apps have fewer than 10 different inconsistent API calls.
It is thus not difficult for developers to perform a one-time manual check.

\begin{figure}[t!]
\begin{adjustbox}{center}
\includegraphics[width=1\textwidth]{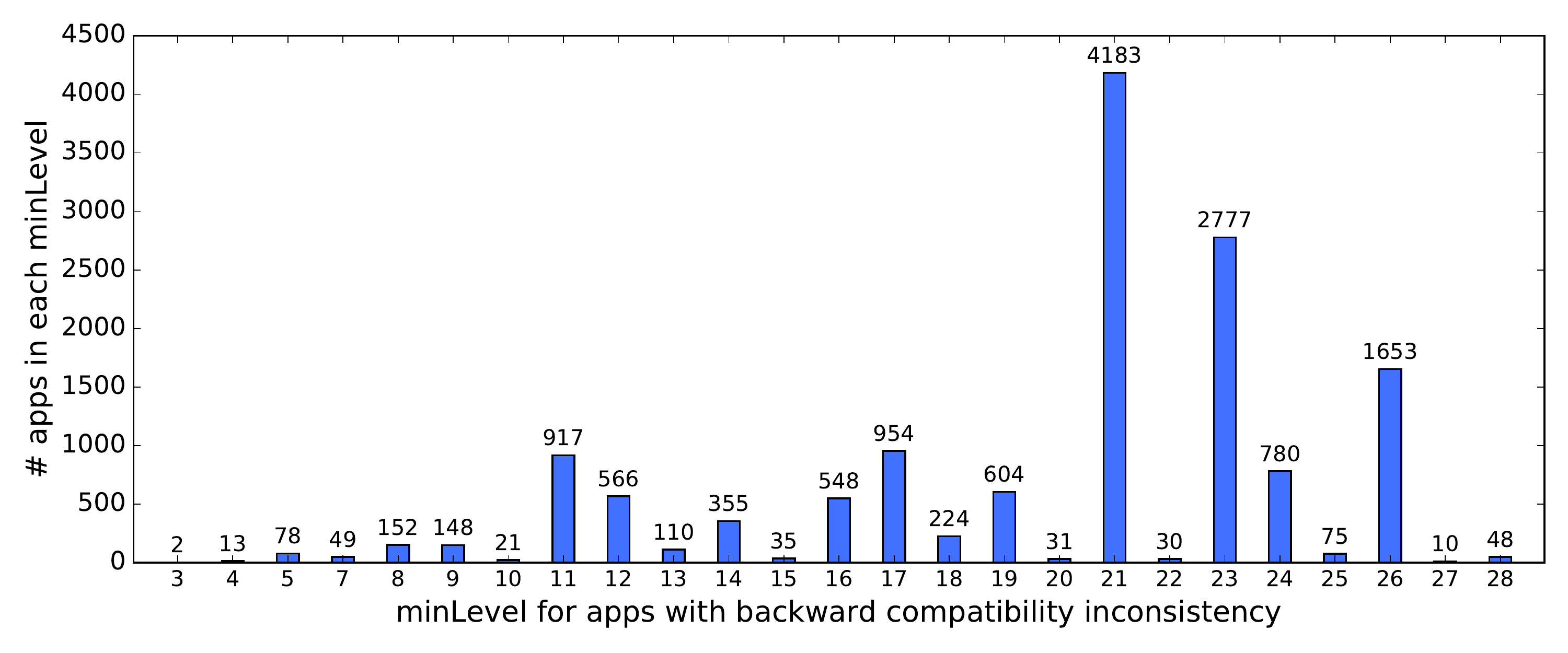}
\end{adjustbox}
\caption{Bar chart of the number of apps in each \minLevel.}
\label{fig:minLevelCrashBar}
\end{figure}

Fortunately, apps with compatibility inconsistency issues could crash \textit{only} on certain Android platforms.
More specifically, they could crash only on versions of platforms between \minSDK and \minLevel, as illustrated earlier in \mysec\ref{sec:sideeffect}.
Therefore, it is necessary to study on which Android platforms those buggy apps could crash, because nowadays some lower versions of Android hold a limited market share, e.g., only 10.7\% for Android lower than 5.0 as of July 2020~\cite{dashboards}.
As a result, even if some apps are buggy with compatibility inconsistency, they cannot trigger the crash on user phones equipped with recent versions of Android.

Since \minLevel is the indicator for maximum versions of Android platforms a buggy app could crash on, we plot a bar chart of \minLevel in \myfig~\ref{fig:minLevelCrashBar} for \red{the} 14,363 app detected with \red{potential} compatibility inconsistency.
We can see that only 2,566 (11.3\% of 22,687) apps could crash on Android 6.0 and above (via counting apps with \minLevel larger than 23), and similarly 1,786 (7.9\%) for Android 7.0 and above.
In other words, most (11,797 out of 14,363) of \red{potentially} buggy apps cannot exhibit their incompatibility bugs on the majority of Android phones that are with 74.8\% market share in July 2020~\cite{dashboards}.
Furthermore, 8,990 out of 14,363 apps could crash only on Android lower than 5.0, which significantly limits the consequences of their incompatibility issues.

\textbf{Finding 2-2:} 
\textit{We find that by employing bytecode search for \texttt{SDK\_INT} checking, our approach can reduce 17.3\% false positives of compatibility inconsistency results.}
As mentioned in \mysec\ref{sec:appAPIcall}, a false positive of compatibility inconsistency could appear if an API call guarded with \texttt{SDK\_INT} checking is not detected.
Here we measure \red{the number of} such false positives \red{that} could be excluded \red{by the} bytecode search.
We find that our search of \texttt{SDK\_INT} checking avoids 3,003 apps from being mistakenly marked with compatibility inconsistency.
Since there are \red{at most} 14,363 apps (i.e., true positives) that could crash when running on lower versions of Android platforms, the percentage of reduced false positives due to bytecode search is \red{at least} 17.3\%.

\begin{figure}[t!]
\begin{adjustbox}{center}
\includegraphics[width=1\textwidth]{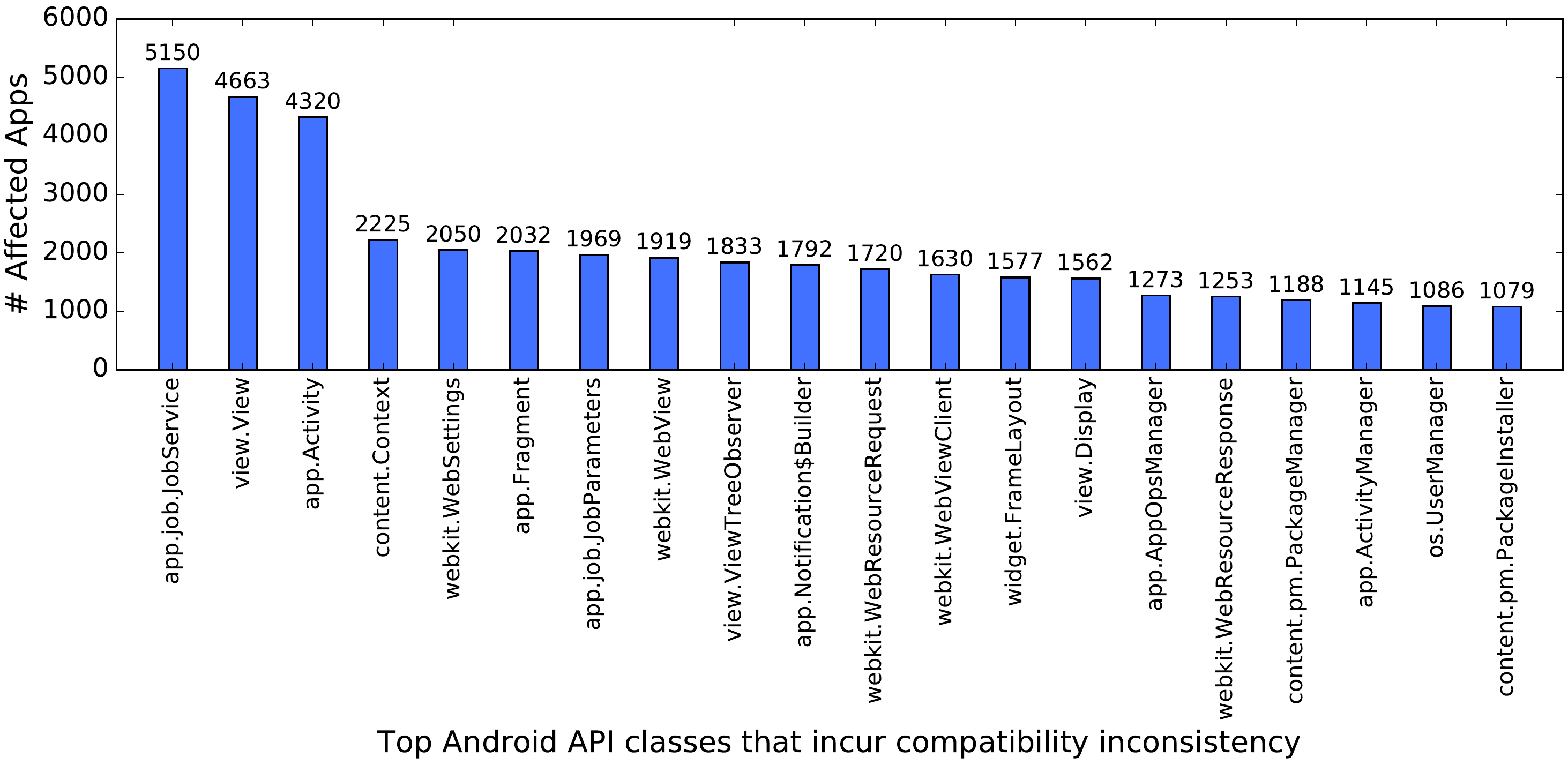}
\end{adjustbox}
\caption{Bar chart of \red{the} top 20 Android API classes (with ``\texttt{android.}'' prefix omitted) that incur compatibility inconsistency in our dataset.}
\label{fig:topCrashClass}
\end{figure}

\textbf{Finding 2-3:} 
\textit{A detailed analysis of Android APIs that incur compatibility inconsistency reveals that some API classes, such as view, webkit, and system manager related classes, are commonly misused.}
We further try to understand the common sources of compatibility inconsistency by analyzing the newly added Android APIs that incur compatibility inconsistency in our dataset.
We find that 6,454 (27.4\% of all 23,542) newly added APIs from 1,138 unique classes cause compatibility inconsistency in at least one app in our dataset.
In particular, 232 (20.4\%) API classes affect more than 100 different apps each, making them the common sources of compatibility inconsistency.
Fortunately, half of API classes only affect fewer than 10 apps each, which suggests that only some portions of API classes \red{are prone to misuses}.

We thus take a closer look at \red{the} top 20 Android API classes that \red{cause} compatibility inconsistency.
As shown in \myfig~\ref{fig:topCrashClass}, all of these classes affect over 1K apps each.
In particular, the \texttt{JobService} class \red{(}introduced in Android 5.0, level 21\red{)} alone could cause compatibility inconsistency in around 5K apps.
Other commonly misused API classes include those related to view (e.g., the \texttt{View}, \texttt{Activity}, \texttt{Context}, and \texttt{Fragment} classes), webkit (e.g., the \texttt{WebSettings} and \texttt{WebView} classes), and system manager (e.g., the \texttt{AppOpsManager} and \texttt{UserManager} classes).
These classes nearly occupy all the top 20 misused ones.

\begin{figure}[t!]
\begin{adjustbox}{center}
\includegraphics[width=0.38\textwidth]{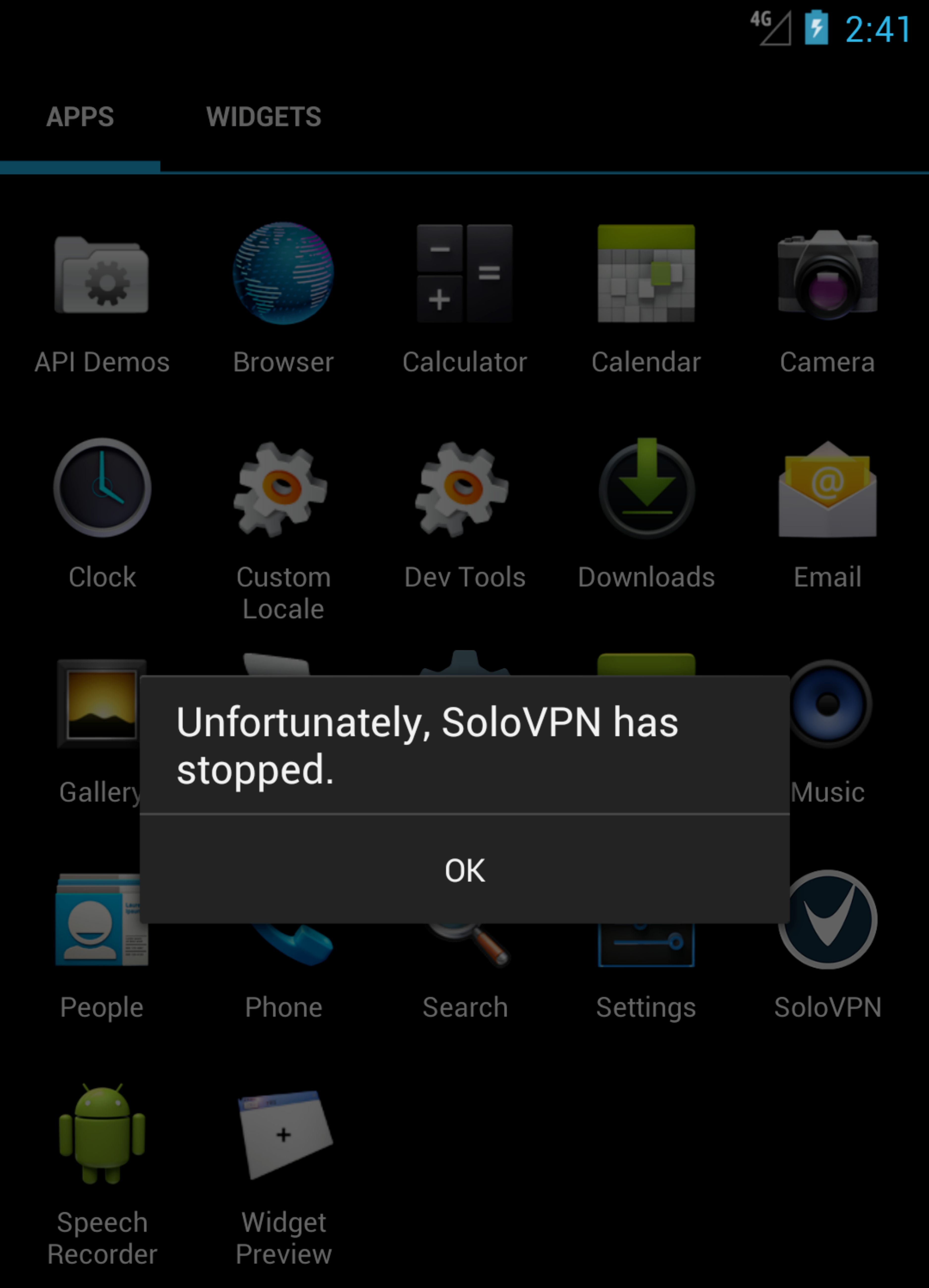}
\end{adjustbox}
\caption{\blue{A case study of the \DSDK issue with incompatibility effect: Solo VPN.}}
\label{fig:crashCase}
\end{figure}

\blue{\textbf{Case study: Solo VPN.}
To demonstrate the impact of incompatibility \DSDK issues, we identify a problematic app in our dataset and try to make it crash at the runtime.
However, it is non-trivial to dynamically achieve this because a crash point may hide deep in certain paths or under certain conditions, which is why the previous work, CiD~\cite{CiD18}, requested developers themselves to help validate their detection results~\cite{CiDReport}. 
To simplify our testing, we intentionally targeted at the VPN apps based on the observation that some \texttt{VpnService} APIs require Android 5.0 at the API level 21.
After testing a few VPN apps in our dataset, we quickly identified a buggy app, Solo VPN (\texttt{co.solovpn}, version: 1.32), which crashed immediately after we clicked the ``Connect'' VPN button on an Android 4.1 device.
\myfig~\ref{fig:crashCase} shows the alert dialog popped up, stating that ``Unfortunately, SoloVPN has stopped''.}


%
%
%
%
%
%

\subsection{RQ3: Inconsistency Results with Security Effect}
\label{sec:rqSECURITY}

In this subsection, we present a total of three findings regarding RQ3.

\textbf{Finding 3-1:}
\textit{Around 2\% apps set an outdated \aimSDK attribute \red{and also invoke} a \red{dangerous} WebView API, making them\red{selves} exploitable by remote code execution.}
As explained in \mysec\ref{sec:effect2}, we measure inconsistency results with the security effect by analyzing each app's \texttt{addJavascriptInterface()} API call and the declared \aimSDK attribute.
In our dataset, 2,791 apps invoke the \texttt{addJavascriptInterface()} API, which suggests that calling this WebView API is necessary in many apps.
However, 484 of them, i.e., 2.1\% of the entire dataset of 22,687 apps, still set an outdated \aimSDK attribute below level 17, making them\red{selves} not only exploitable on Android lower than 4.2 but also vulnerable on higher versions of Android platforms.
This could be avoided if their \aimSDK values \red{are} updated.


\textbf{Finding 3-2:} 
\textit{Our bytecode search of \texttt{addJavascriptInterface()} invocation helps reduce 12.2\% false positives.}
Recall \red{from} \mysec\ref{sec:appAPIcall} \red{that} we perform bytecode search to check whether an \texttt{addJavascriptInterface()} API call is invoked by a valid class.
We find that without such checking, 551 apps can be detected with vulnerable combination of \texttt{addJavascriptInterface()} and \aimSDK.
In other words, our search of \texttt{addJavascriptInterface()} invocation avoids 67 (551 - 484) apps from being mistakenly marked with security inconsistency.
Hence, we conclude that our bytecode search reduces 12.2\% false positives in the context of \texttt{addJavascriptInterface()}.

\begin{table}[h!]
  \caption{\red{The top} five library classes that introduce \texttt{addJavascriptInterface()} API call in vulnerable apps and the number of apps affected.}
\begin{adjustbox}{center}
\begin{tabular}{  c| c}

\hline
Library Class & \# Vulnerable Apps \tabularnewline
\hline
\hline

Lcom/flurry/android/CatalogActivity;  & 41 \tabularnewline
\hline

Lcom/openfeint/internal/ui/NativeBrowser;  & 30 \tabularnewline
\hline

Lcom/doodlemobile/gamecenter/moregames/MoreGamesActivity;  & 19 \tabularnewline
\hline

Lcom/gau/go/launcherex/theme/classic/FullScreenAdWebPage;  & 17 \tabularnewline
\hline

Lcom/amazon/ags/html5/overlay/GameCircleUserInterface;  & 13 \tabularnewline
\hline

\end{tabular}
\end{adjustbox}
\label{tab:vulnerableLibrary}
\end{table}

\textbf{Finding 3-3:}
\textit{Around half of \red{the} vulnerable apps invoke the \texttt{addJavascript} \texttt{Interface()} API \red{only} because of their embedded third-party libraries.}
Our approach \red{can also determine} whether the \texttt{addJavascriptInterface()} API is invoked by app's own code or embedded by a third-party library.
It turns out that 214 (44.2\%) of 484 vulnerable apps invoke \texttt{addJavascriptInterface()} \red{only} because of their embedded third-party libraries.
In particular, five libraries affect at least 10 vulnerable apps each.
Table~\ref{tab:vulnerableLibrary} lists their class names and the number of apps affected.
We can see that the popular Yahoo Flurry SDK~\cite{FlurrySDK} and OpenFeint Game SDK~\cite{OpenFeint} cause some apps with outdated \aimSDK vulnerable.

This finding gives two implications.
\red{First}, developers must check whether a third-party library invokes some vulnerable APIs before embedding it into apps.
\red{Second}, library producers also need to ensure \red{that} certain dangerous APIs are invoked only in safe versions of Android platforms, because a library can be used in any app with all kinds of \aimSDK values.

\begin{figure}[t!]
\begin{adjustbox}{center}
\includegraphics[width=0.7\textwidth]{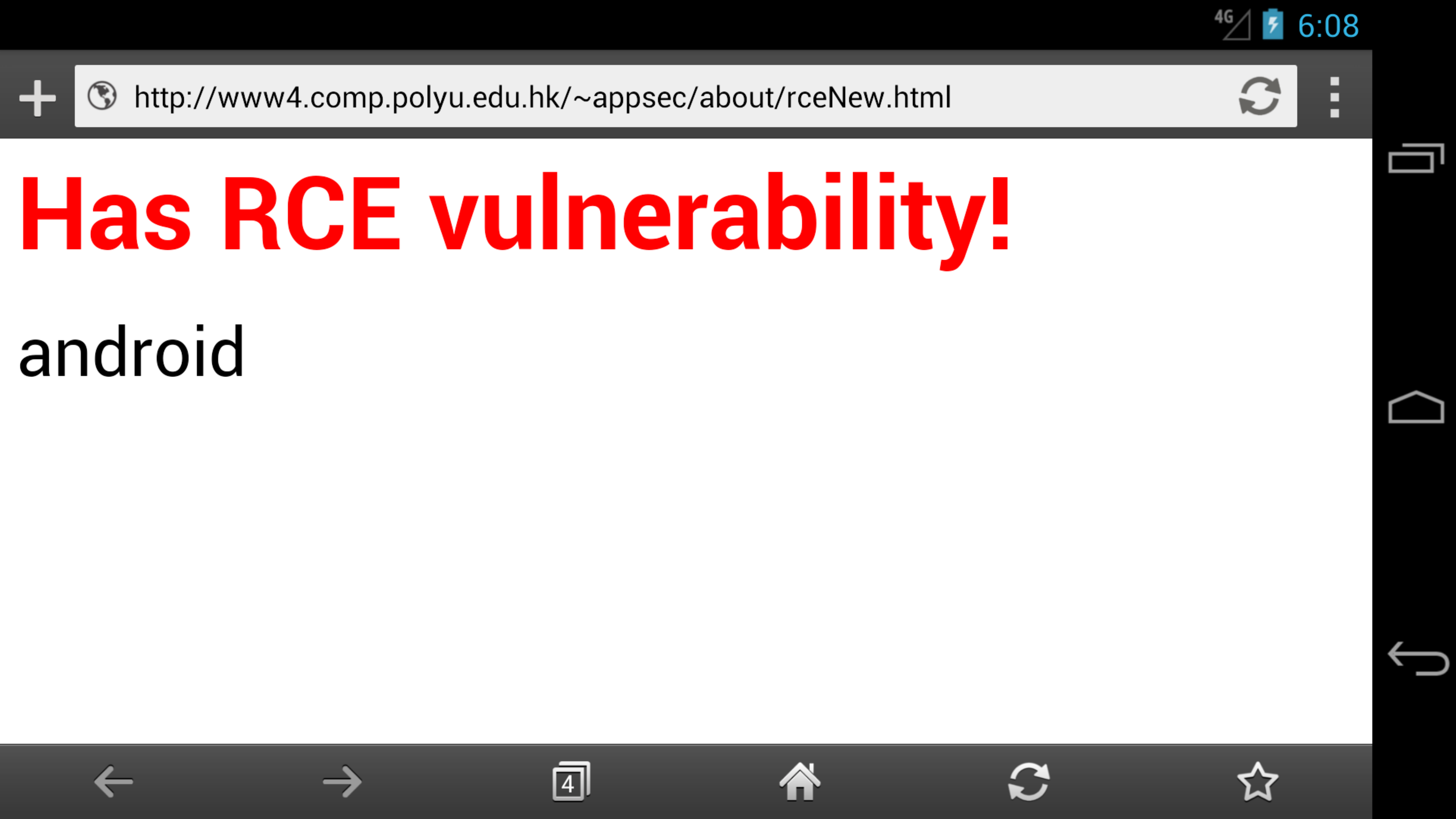}
\end{adjustbox}
\caption{\blue{A case study of the \DSDK issue with security effect: Exsoul Browser.}}
\label{fig:vulnerableCase}
\end{figure}

\blue{\textbf{Case study: Exsoul Browser.}
To demonstrate the impact of insecure \DSDK issues, we try to exploit a problematic app in our dataset.
To exploit \texttt{addJavascript} \texttt{Interface} vulnerabilities, an adversary needs to inject a piece of malicious Javascript code into a vulnerable WebView-based interface in the victim app.
He or she could achieve this \red{by either} intercepting the HTTP traffic via a Man-In-The-Middle proxy or tricking victim users to directly browse a malicious website.
We chose the second more convenient way and directly targeted at the browser apps in our dataset for tests.
There was only one browser app, Exsoul Browser (\texttt{com.exsoul}), reported with \DSDK security problems.
We used it to browse a demonstration exploit website that we prepared before, \url{http://www4.comp.polyu.edu.hk/~appsec/about/rceNew.html}, which would output ``Has RCE Vulnerability'' if the tested browsing interface is vulnerable. 
As shown in \myfig~\ref{fig:vulnerableCase}, we successfully validate the \path{addJavascriptInterface} vulnerability in Exsoul Browser on an Android 4.1 device.
We also find that Exsoul Browser exposed a Javascript interface named ``android'', which allows a malicious website to execute arbitrary commands by simply invoking this Javascript code: \path{android.getClass().forName("java.lang.Runtime").getMethod("getRuntime",null).invoke(null,null).exec(cmdArgs)}.}

\subsection{RQ4: Performance Metrics of Our Approach}
\label{sec:rqPERFORMANCE}

In this section, we evaluate performance metrics of our approach to answer RQ4.

\begin{figure}[t!]
\begin{adjustbox}{center}
\includegraphics[width=0.6\textwidth]{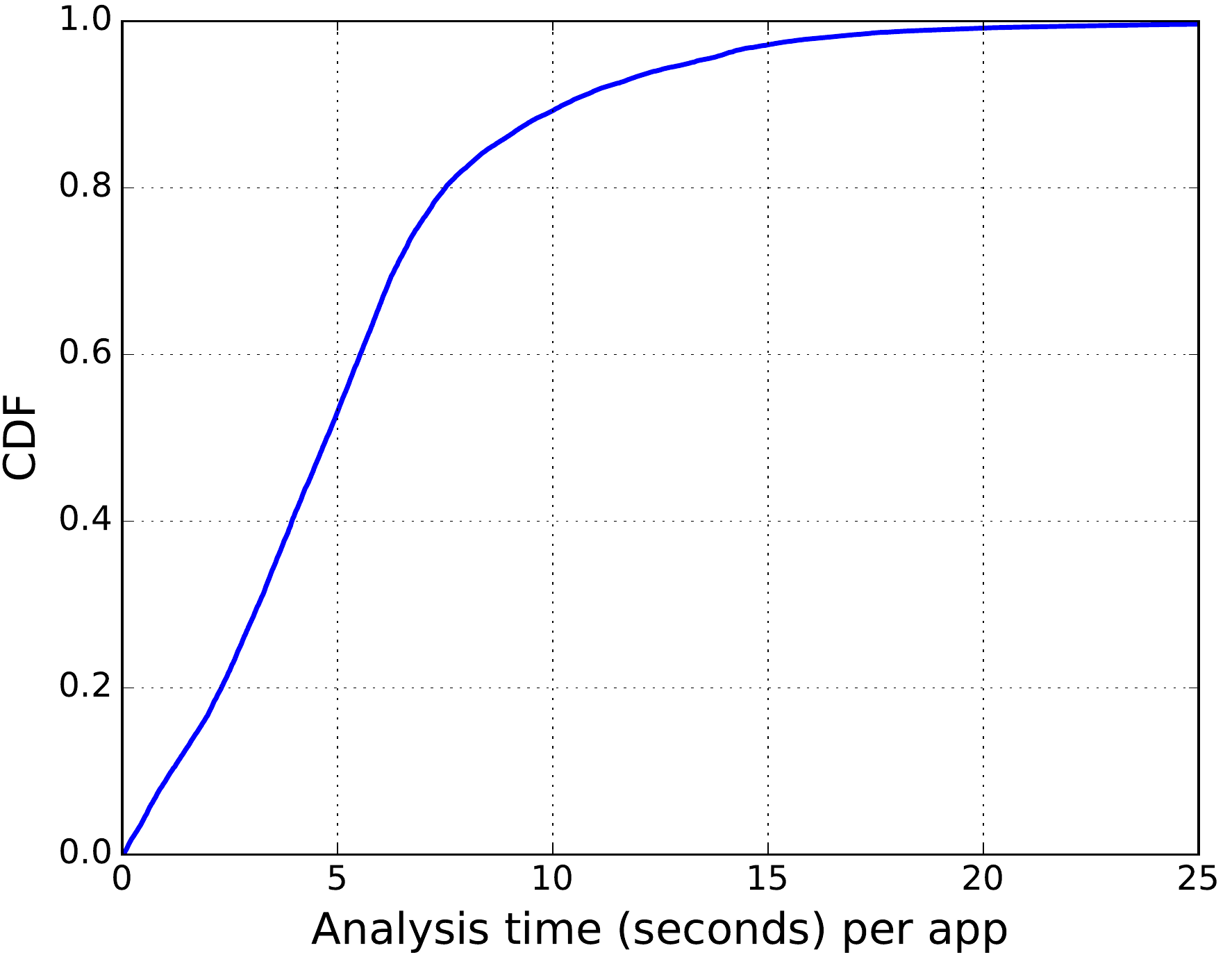}
\end{adjustbox}
\caption{CDF plot of the \red{amount of} time \red{required} for our approach to analyze each app.}
\label{fig:runningTime}
\end{figure}

\textbf{Finding 4-1:}
\textit{Our approach achieves good scalability with an average time of 5.39s and the analysis time of 90\% apps in less than 10 seconds. This makes our approach suitable for online vetting.}
In \myfig~\ref{fig:runningTime}, we present CDF plot of the \red{amount of time required for our approach} to analyze each app.
We can see that more than 50\% apps can be analyzed in less than five seconds each, with the median time of 4.75s.
The average analysis time of all the 22,687 apps is only 5.39s.
These results indicate that our approach achieves good scalability\red{, therefore} suitable for online vetting. 
App markets can deploy our approach to timely notify developers the \DSDK inconsistency in their apps.

In contrast, dataflow-based approaches~\cite{CiD18}~\cite{IctApiFinder18} suffer from the scalability problem.
Specifically, CiD~\cite{CiD18} failed to analyze 387 apps (out of a dataset of 2,000 apps) due to timeouts and bugs.
This 19.4\% timeout or failure rate makes it infeasible for online vetting, let alone performance statistics were also not clear for those successfully analyzed.
On the other hand, IctApiFinder~\cite{IctApiFinder18} takes 3 minutes and 45 seconds to analyze only an app of 8MB (the app is available via historical versions on \url{https://f-droid.org/en/packages/com.nextcloud.client/}), a size much smaller than the average size (25MB) of our dataset.
This suggests that IctApiFinder is impractical to perform online vetting of a modern app dataset from Google Play (all apps evaluated by IctApiFinder were open-source apps from the F-Droid website).

\begin{figure}[t!]
\begin{adjustbox}{center}
\includegraphics[width=0.6\textwidth]{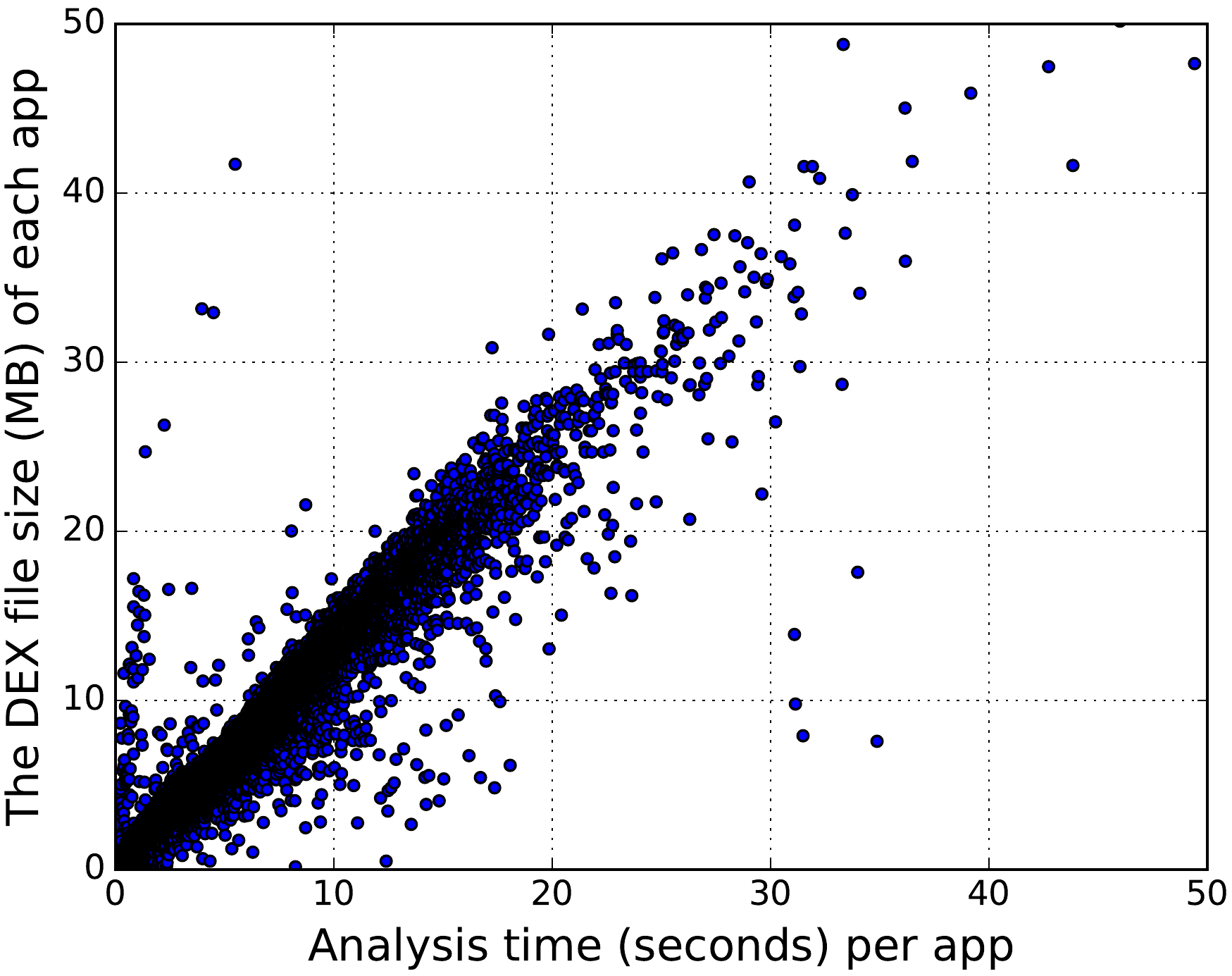}
\end{adjustbox}
\caption{Scatter plot of the relationship between analysis time and DEX size.}
\label{fig:timeDEXsize}
\end{figure}

\textbf{Finding 4-2:}
\textit{A further correlation analysis between analysis time and app size shows that the performance of our approach is approximately \red{in} a linear relationship with DEX file size of the app.}
We \red{find} that the performance of our approach is always under control regardless of app size.
This can be evaluated by performing a correlation analysis between analysis time and app size.
In \myfig~\ref{fig:timeDEXsize}, we draw a scatter plot of the relationship between analysis time and the size of DEX file of the app (APK file contains both bytecode and resource files while DEX file is only for bytecode).
According to this figure, \red{the} analysis time and DEX file size \red{are} approximately \red{in} a linear relationship, \red{at the} rate of around 30 seconds for a 40MB DEX file (note that we count the file size of multiple DEX files if any).
There are some outliers of small apps with more analysis time (e.g., five apps under 20MB exceeding 30s), which is largely because these apps involve much more vulnerable API calls to search.
On the other hand, the outliers of large apps with less analysis time is due to unused third-party libraries embedded.
Overall, the linear relationship between analysis time and app size indicates that our approach can achieve good performance even with large apps.

\subsection{\blue{RQ5: The Updatability of The Buggy Apps}}
\label{sec:rqUPDATE}

In this subsection, we continue to understand the updatability of apps that were measured with \DSDK issues in our dataset, i.e., whether they are still maintained by their developers.
This is important because compared with the updatable apps that could eventually address their \DSDK issues via the app updates, outdated apps have no maintainers to periodically update and fix their \DSDK problems.
To study to what extent this problem is, we use 8,359 unique apps (8,019 incompatible apps and 484 vulnerable apps) that were reported with potential \DSDK problems in \mysec\ref{sec:rqCRASH} and \ref{sec:rqSECURITY} for the analysis.
Since our dataset was crawled in November 2018, we collected the latest release date of those buggy apps on Google Play in early December 2019.
We believe that this one-year time frame allows us to test the app updatability by analyzing whether apps have been updated in 2019 or not. 
We show our finding in the next paragraph.

\begin{figure}[t!]
\begin{adjustbox}{center}
\includegraphics[width=0.75\textwidth]{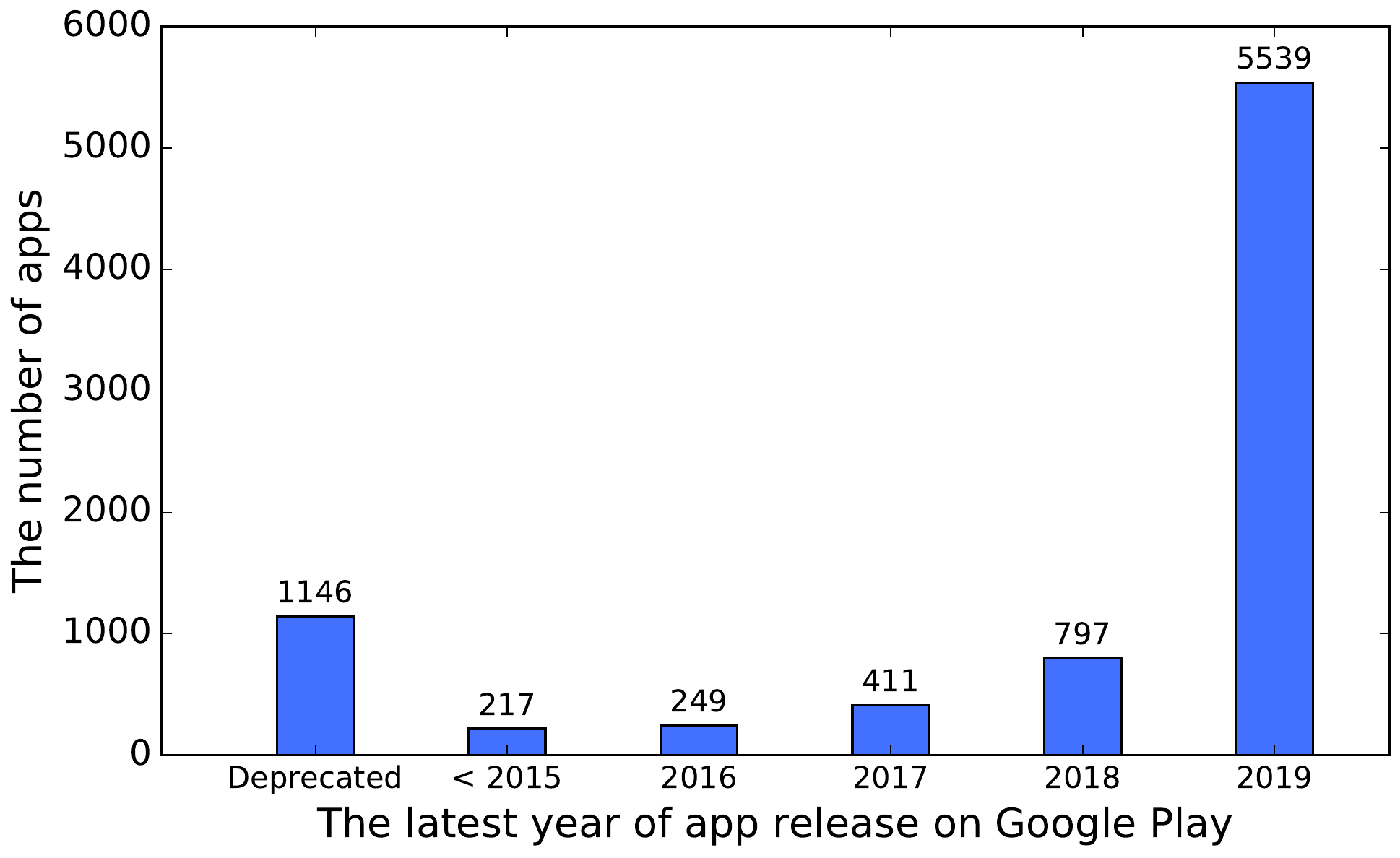}
\end{adjustbox}
\caption{Bar chart of the distribution of apps that were measured with \DSDK issues in our dataset and their latest release years on Google Play.}
\label{fig:withYearBar}
\end{figure}

\textbf{Finding 5:}
\textit{Around 20\% of the 8,359 buggy apps were never updated in 2019, and 13.7\% have been deprecated from Google Play, causing a total of 33.7\% apps outdated.}
\myfig~\ref{fig:withYearBar} shows a bar chart of the distribution of apps that were measured with \DSDK issues in our dataset and their latest release years on Google Play.
According to this figure, 5,539 (66.3\%) apps have been updated at least once in 2019, which allows their developers to upgrade \DSDK versions to fix their \DSDK problems.
However, there are still one third of the measured apps not updatable.
Specifically, the latest release years of 1,674 (20\%) apps have been 2018, 2017, 2016, and even before 2015.
Besides these ``old'' apps, we find that 1,146 (13.7\%) apps are even deprecated from Google Play for various reasons (e.g., being taken down by developer themselves or violating the advertisement policy on Google Play).
No matter for what reasons, they are no longer on Google Play due to no further maintenance, whereas their previously downloaded versions could still be in user phones.
Both old and deprecated apps incur a large number of outdated apps in the wild, with a total of 33.7\% in our dataset.
Therefore, it is worthwhile for researchers to further develop techniques for automatically fixing \DSDK issues in those outdated apps.

\section{\blue{Implications}}
\label{sec:implication}

In this section, we further present two implications on the qualitative analysis of identified \DSDK problems and actionable countermeasures for developers.

\textbf{Implication 1:}
\textit{Android's original design of the \texttt{DSDK} mechanism, despite the good intention, does not satisfy the expectation of developers' real usage.}
One major problem is that it is difficult to evolve the \texttt{DSDK} versions correctly when apps are updated with new or deprecated APIs.
The original \texttt{DSDK} design is a \textit{static} mechanism, and there was no automatic mechanism to \textit{dynamically} update the outdated \texttt{DSDK} versions.
However, quite a number of apps are updated frequently, e.g., 1,448 of the top 10,713 apps studied in 2014 were updated on a bi-weekly basis or even more frequently~\cite{FreshApp15}.
In this way, it is challenging for developers to maintain the \texttt{DSDK} versions while they are already busy with the functionality update.
Moreover, the \texttt{addJavascriptInterface()} vulnerabilities reported in \mysec\ref{sec:rqSECURITY} indicate that there is a semantic gap between the \texttt{targetSdkVersion} design and developers' understanding.
Indeed, it is somehow confusing that lower versions of API behaviors would be used even when an app is running on a higher version of the Android platform (see \mysec\ref{sec:backg}).
To our knowledge, this is not the first case where a misunderstanding between Android's design and developers' knowledge happens.
Another notable example is that Android once by default exported all content provider components that have no \texttt{android:exported} attribute defined, which caused a large number of vulnerable apps~\cite{ContentScope13} since developers did not expect their content provider components to be exported.

It is worth noting that this implication is only our plausible conjecture.
Based on the factual analysis results reported in \mysec\ref{sec:evaluate}, other conjectures could also be possible.
However, no matter what the causes are from the developers' perspective, the consequences remain the same and significant.

\textbf{Implication 2:}
\textit{To mitigate the \DSDK problems, the Android community could take countermeasures from different levels.}
We list the following three actionable countermeasures that can be adopted by different stakeholders: 
\begin{itemize}
  \item Google Android could provide better IDE (integrated development environment) to help developers check \DSDK versions before uploading their apps to the markets.
    Such checking is ideally automatic and should launch whenever there are new changes in apps.
    We have seen a good trend in the recent Android Studio IDE, which performs more user-friendly \DSDK checking than its predecessor, i.e., the Android Lint plugin in Eclipse.

  \item The app markets can deploy our approach to perform a quick and mandatory checking of each \red{uploaded} app.
    The suspicious \DSDK conflicts and recommendations \red{need} to be either approved or dismissed.
    In this way, we can guarantee that developers are at least aware of potential \DSDK problems in their apps.

  \item As the last line of defense, end-user Android devices can dynamically upgrade \DSDK versions in victim apps or enforce mandatory access control~\cite{SCLib18} so that they are no longer incompatible or vulnerable at the operating system level.
    This is especially important for the apps no longer maintained (see \mysec\ref{sec:rqUPDATE}).
\end{itemize}

\section{Threats To Validity}
\label{sec:discuss}

In this section, we \blue{summarize some major} threats to the validity of our study.

Firstly, \red{same as} typical Android static analysis, our approach does not handle Java reflection, dynamic code loading, native code, and complicated code obfuscation.
However, some apps may employ these mechanisms to access certain Android APIs.
If \red{one} such API call has inconsistency issues, a false negative would appear.
Since these code protection mechanisms are usually used in malware, our statistical results of popular apps will be less affected and we \red{will consider} these mechanisms \red{in} our future work.

Secondly, although our bytecode search in \mysec\ref{sec:appAPIcall} has minimized false positives caused by \texttt{VERSION.SDK\_INT} checking and uninvoked third-party libraries, it is theoretically less accurate than dataflow-based approaches.
\blue{Fortunately, in our deployment model, we can rely on developers to manually check and correct inconsistency reported by our approach.}
\blue{Moreover}, as evi\red{de}nced in \mysec\ref{sec:rqCRASH}, \blue{the manual effort required in such checking is also limited ---} around 80\% apps are reported with fewer than ten inconsistent API calls each, which is manageable for developers to perform a one-time manual check.
\blue{Due to this limitation, the measurement results reported in this paper represent an upper bound of all potential \DSDK problems (under the condition that the common analysis difficulties above are not considered).
This satisfies our objective of conducting a comprehensive \DSDK study, whereas it is not suitable for bug detection.}

Thirdly, the consistency detection in this paper focuses on changed APIs, but there are also added and removed Java/Android fields \red{during the} SDK evolution.
To build the mapping between fields and SDK versions, we found that we can leverage the same document analysis method in \mysec\ref{sec:DocAnalysis}, because the \texttt{api-versions.xml} file also records added, removed, and deprecated fields in all Android classes.
By inputting this mapping to our app analysis, we can extend our consistency detection to evolved Android fields as well in \red{our} future \red{work}.

Lastly, although we have updated the original 2015 dataset with a recent dataset crawled in November 2018 and further checked its updatability in December 2019, we are not able to keep updating it. 
As a result, the findings reported in this paper may not represent the latest scenarios.
We invite other researchers to replicate our findings on more recent datasets.

\section{Related Work}
\label{sec:related}

In this section, we summarize some related research on declared SDK versions, Android APIs, and Android app static analysis.

\subsection{Research on Declared SDK Versions}
To the best of our knowledge, there were no systematic studies on declared SDK versions previously, \red{except for} some specific studies on the \aimSDK or \minSDK attributes in different scenarios.
Notably, Wu and Chang~\cite{FileCross14} showed that due to using outdated \aimSDK attributes, many Android browser apps were vulnerable to \texttt{file://} vulnerabilities.
They further demonstrated more security consequences caused by outdated \aimSDK attributes~\cite{MoST15}.
Following this line of research, Mutchler et al.~\cite{TargetFragment16} conducted a large-scale measurement of multiple vulnerabilities that are affected by the fragmented \aimSDK attributes.
Wei et al.~\cite{TameFragment16} also studied Android fragmentation with the focus on compatibility issues.
In particular, our \red{preliminary} conference version of this work~\cite{WASA17} has \red{motivated} three recent follow-up works~\cite{CiD18}~\cite{IctApiFinder18}~\cite{ACRYL19} on detecting compatibility issues caused by inappropriate \minSDK attributes.
Compared to all these works, our study is the first systematic work on measuring all kinds of \DSDK versions and their (in)consistency with API calls.

\subsection{Android API Studies}
Besides \DSDK and fragmentation, our paper is also related to prior studies on Android APIs or SDKs.
Among these studies, the work performed by McDonnell et al.~\cite{ICSM13} is the closest to our paper.
They also studied the Android API evolution but \red{focused on} how client apps follow Android API changes\red{. In contrast,} \red{our} focus \red{is} the consistency between apps' \DSDK and API calls.
Other related works have studied the \red{correlation} between apps' API change and their success~\cite{AppSuccess13}, the deprecated API usage in Java-based systems~\cite{DeprecatedAPI16}, the inaccessible APIs in Android framework and their usage in third-party apps~\cite{Inaccessible16}\red{,} and the Android Alarm API usage and their impacts to network latency~\cite{AppAlarm16}.
In particular, the work \red{conducted by Almeida et al.}~\cite{AppAlarm16} analyzed \aimSDK in \red{the} apps that invoke Alarm APIs.
Additionally, several security papers analyzed the mappings between Android APIs and their permissions~\cite{Stowaway11}~\cite{PScout12}~\cite{Axplorer16}.

\subsection{Android App Static Analysis}
A large number of Android studies have leveraged static analysis in many applications over past years.
The major methodology can be roughly classified into control-flow based reachability analysis and dataflow-based taint analysis.
For the reachability analysis, RiskRanker~\cite{RiskRanker12} and Woodpecker~\cite{Woodpecker12} are two pioneer representative works in the domains of malware detection and vulnerability discovery, respectively.
They tested the reachability from entry points to sink APIs.
In contrast, more prior works employed dataflow analysis to taint the propagation flows of an interested data variable.
FlowDroid~\cite{FlowDroid14}, Amandroid~\cite{Amandroid14}, DroidSafe~\cite{DroidSafe15}, and HornDroid~\cite{HornDroid16} are representative works in this research direction.
In particular, FlowDroid and Amandroid have been used or customized in many follow-up static analysis tools (e.g.,~\cite{AppContext15}~\cite{MudFlow15}~\cite{DomainSocket16}~\cite{OpenPort17}~\cite{IctApiFinder18}).
One common thing between reachability analysis and dataflow analysis is that they both require to generate an app call graph, the precision of which affects the entire analysis accuracy.
However, generating a high-precision call graph requires expensive pointer analysis~\cite{Amandroid14}, and the scalability concern is why we proposed lightweight bytecode search for our online vetting of API-SDK inconsistency in this paper.

\section{Conclusion and Future Work}
\label{sec:conclude}

In this paper, we \red{conducted} a systematic study of declared SDK versions in Android apps, a modern software mechanism that received little attention. 
We measured the current practice of declared SDK versions or \DSDK versions in a large set of 22,687 modern apps and the inconsistency between \DSDK versions and their host apps' API calls.
To facilitate the analysis that can be \red{readily} deployed by app markets for online vetting, we proposed a robust and scalable approach that operates on the Android bytecode level and employs a lightweight bytecode search for app analysis.
We have obtained some interesting new findings, including (i) 4.76\% apps do not claim the targeted \DSDK versions, although this percentage has significantly dropped over recent three years, (ii) \red{around 35\%} apps under-set the minimum \DSDK versions and could incur runtime crashes, but fortunately, only 11.3\% apps could crash on Android 6.0 and above, and (iii) around 2\% apps, due to under-claiming the targeted \DSDK versions, are potentially exploitable by remote code execution, and half of them invoke the vulnerable API via embedded third-party libraries.
In the future, we plan to conduct more \DSDK case studies and report buggy cases to app developers and markets for fixes, and further improve our approach to mitigate some threats to validity.

\begin{acknowledgements}
We thank editors and all the reviewers for their valuable comments and helpful suggestions.
This work is partially supported by a direct grant (ref. no. 4055127) from The Chinese University of Hong Kong.
\end{acknowledgements}

\bibliographystyle{spmpsci}      
\bibliography{main}

\end{document}